\documentclass[12pt]{article}
\usepackage[a4paper,margin=1in]{geometry}
\usepackage{authblk} % For \author[1], \affil, etc.
\usepackage{graphicx} % Required for inserting images
\usepackage{subcaption} % For subfigures + individual labels
\usepackage{subfiles} % Allow separate compilation of additions/chapters via \subfile{}
\usepackage{pdflscape} % for landscape pages (Figure 8)
\usepackage{subscript} % For \textsubscript
\usepackage{amsmath} % For \text in math mode
\usepackage{setspace}
\usepackage[super,sort&compress]{natbib} % ACS-style numeric superscript citations (ordered by appearance)
\usepackage{tikz} % for drawing overlays on figures (e.g., Fig. 11b diagonal)

\usepackage{hyperref}
\usepackage{float} % provides the [H] float placement specifier
\usepackage{placeins} % provides \FloatBarrier to keep floats within sections

\usepackage{titlesec}
\titleclass{\subsubsubsection}{straight}[\subsubsection]
\newcounter{subsubsubsection}[subsubsection]
\renewcommand\thesubsubsubsection{\thesubsubsection.\arabic{subsubsubsection}}
\titleformat{\subsubsubsection}{\normalfont\normalsize\bfseries}{\thesubsubsubsection}{1em}{}
\titlespacing*{\subsubsubsection}{0pt}{3.25ex plus 1ex minus .2ex}{1.5ex plus .2ex}
\title{A Novel Method to Determine Total Oxidant Concentration Produced by Non-Thermal Plasma Based on Image Processing and Machine Learning}
\author[2]{Mirkan Emir Sancak}
\author[2]{Unal Sen}
\author[1,*]{Ulker Diler Keris-Sen}

\affil[1]{Institute of Earth and Marine Sciences, Gebze Technical University, TR-41400, Kocaeli, Turkey}
\affil[2]{Department of Environmental Engineering, Gebze Technical University, TR-41400, Kocaeli, Turkey}

\affil[*]{Corresponding Author: Tel.: +90 (262) 6053163 \\ E-mail: udkeris@gtu.edu.tr}
\date{}

\begin{document}

\maketitle

\begin{abstract}
    Accurate determination of total oxidant concentration ($[\text{Ox}]_{\text{tot}}$) in non-thermal plasma (NTP)-treated aqueous systems remains a critical challenge due to the transient nature of reactive oxygen and nitrogen species and the subjectivity of conventional titration methods used for $[\text{Ox}]_{\text{tot}}$ determination. This study introduces a novel, color-based computer analysis (CBCA) method that integrates advanced image processing with machine learning (ML) to quantify colorimetric shifts in potassium iodide (KI) solutions during oxidation. First, a custom-built visual data acquisition system captured high-resolution video of the color transitions in a KI solution during oxidation with an NTP system. The change in $[\text{Ox}]_{\text{tot}}$ during the experiments was monitored with a standard titrimetric method. Second, the captured frames were processed using a robust image processing pipeline to extract RGB, HSV, and Lab color features. The extracted features were statistically evaluated and the results revealed strong linear correlations with the measured $[\text{Ox}]_{\text{tot}}$ values, particularly in the saturation (HSV), a and b (Lab), and blue (RGB) features. Subsequently, the $[\text{Ox}]_{\text{tot}}$ measurements and the extracted color features were used to train and validate five ML models. Among them, linear regression and gradient boosting models achieved the highest predictive accuracy ($R^2 > 0.990$). Reducing the input from nine to four and three key features preserved high predictive accuracy while improving computational efficiency, with ensemble models---especially gradient boosting---remaining robust after feature reduction. Finally, comparison of the model predictions with real titration measurements revealed that the CBCA system successfully predicts the $[\text{Ox}]_{\text{tot}}$ in KI solution with high accuracy ($R^2 > 0.998$) even with reduced number of features. 
\end{abstract}

\textit{\noindent\textbf{Keywords:} Reactive oxygen and nitrogen species, Total oxidant, Machine learning, Image processing, Iodometric titration method
}

\textit{\noindent\textbf{ORCID:} Mirkan Emir Sancak (0000-0001-7237-2047), Unal Sen (0000-0003-1884-9152), Ulker Diler Keris-Sen (0000-0002-8354-1640)
}
\section{Introduction}
The non-thermal plasma systems (NTP) are one of the advanced oxidation processes used in water, wastewater, and gas treatment for decades. In these systems, various reactive oxygen and nitrogen species (RONS) are produced such as ozone (O$_3$), hydroxyl radicals (OH$\cdot$), hydrogen peroxide (H$_2$O$_2$), nitric oxide (NO$\cdot$), peroxynitrites, peroxynitrates, etc., and their possible intermediates, as a function of plasma source, discharge properties, and feed gas, etc. \cite{Angelina2025ATreatment}. The lifetime of these species changes from nanosecond to minute, depending on the environmental condition and presence of the scavengers, \cite{Phaniendra2015FreeDiseases} such as OH$\cdot$ radicals' lifetime is 0.1 ns but, O$_3$ stable from 1 to 20 minutes depending on the composition of water \cite{Deng2019ATest}. There are specific methods for individual quantitative analysis of a short-lived species in the water. For example, OH$\cdot$ radicals are measured by modified Griess test \cite{Deng2019ATest}, or visual color detection with methylene blue dye \cite{Satoh2007MethyleneSolution}. On the other hand, relatively long-lived species, such as O$_3$, are measured by the indigo method. Besides, H$_2$O$_2$ concentration in the aqueous phase is measured by ceric sulfate \cite{Furman1930ASulfate}, potassium permanganate, or potassium iodide solutions \cite{Schumb1994H202Titration}.

RONS represents the total oxidants of any NTP system. Therefore, measuring the concentration of RONS is important to determine the efficiency of the NTP system performance. In previous studies, the applicability of the iodometric method (KI method) has been reported for the quantitative determination of total oxidant concentration ($[\text{Ox}]_{\text{tot}}$) \cite{Wang2022CoolingDetermination, Deadman2017APeroxide}. Besides, determination of the $[\text{Ox}]_{\text{tot}}$ in any advanced oxidation process and/or chemical oxidation system can also be provided by KI method. For instance, Liang and He (2018) studied the in-situ dual chemical oxidation system for soil and groundwater remediation using two oxidizing agents, sodium persulfate and hydrogen peroxide (H$_2$O$_2$). They showed that the KI method was acceptable for determination of $[\text{Ox}]_{\text{tot}}$ \cite{Liang2018ASystem}. Another study investigated the rapid and selective determination of the total oxidant capacity of the peroxodisulfate, peroxomonosulfate, and H$_2$O$_2$-containing samples. They reported that the KI method is not selective for the oxidant but applicable for the determination of $[\text{Ox}]_{\text{tot}}$ \cite{Deadman2017APeroxide}.

In the KI method, the thermodynamic instability of iodide (I$^-$) ion reacts with any oxidants to form iodine (I$_2$) \cite{MirelladaSilva2024RemediationOxidants, MindiaAli2024RecentStudies}, so the initially colorless KI solution turns a yellow-to-brown color change depending on the presence and concentration of the oxidants. The concentration of the total oxidant is conventionally calculated by back titration of the KI solution with sodium thiosulfate Standard Methods 2350-E \cite{1998StandardWastewater} which the KI solution can be used repeatedly \cite{Sancak2025ReusabilityDegradation} or spectrophotometrically \cite{Wang2022CoolingDetermination}. The accuracy of the back-titration step is mostly dependent on the operator’s visual perception. The perception and acquisition of an image can vary depending on the observer, particularly due to variations in individual eye sensitivity \cite{Tulver2019ThePerception}. Human vision, influenced by factors such as operator fatigue, ambient lighting, and perceptual biases, introduces variability in visual data interpretation. These inconsistencies can lead to deviations in visual analysis results, particularly in applications requiring precise quantitative assessments, such as colorimetric measurements or defect detection \cite{Bosten2022DoPerception}.

Image processing techniques can be integrated with ML for the determination of various color-based substances to eliminate observer-dependent discrepancies \cite{Wolin1998TheEvaluation, Ding2021ComparisonSystems, Gastaldo2005ObjectiveNetworks}. The extraction of color binaries is crucial in colorimetric titration for enhancing precision and accuracy by minimizing external factors like ambient light and indicator dichroism, ensuring precise endpoint detection, and limited human perception \cite{Huang2021ATitrations}. To date, several researchers have investigated only-image processing \cite{Puangbanlang2019ASamples} or image processing and ML combination, highlighting its potential in diverse analytical applications. In the literature, comprehensive experiments have been conducted to evaluate ML-based colorimetric analysis of paper-based-analytical-devices for pesticide residue detection, employing various models such as Linear Regression (LiR), Support Vector Machine (SVM), Random Forest Regressor (RFR), Artificial Neural Network (ANN), and color spaces like RGB, HSV, Lab under diverse lighting, camera, and user conditions \cite{Khanal2021Machine-Learning-AssistedDevices}.

Machine learning significantly enhances the accuracy and efficiency of concentration determination in colorimetric analysis by improving data analysis and determination capabilities, even under varying conditions. Among these ML models such as ANN, SVM, logistic regression, and RFR have demonstrated remarkable improvements in determination accuracy for colorimetric analysis by analyzing images captured under diverse lighting conditions and using various devices \cite{Khanal2021Machine-Learning-AssistedDevices}. Khanal et al. showed that ANN with the Lab color space performed best for food dye analysis, while SVM with Lab excelled in pesticide analysis. The Lab color space was more accurate and consistent than other spaces, and ANN and SVM demonstrated the highest performance and generalization capabilities \cite{Khanal2021Machine-Learning-AssistedDevices, KhanalColorimetricLearning}. Lee et al. demonstrated the effectiveness of the RFR model, achieving excellent performance in predicting pH levels and glucose concentrations using colorimetric paper sensors \cite{Lee2022ASensors}. Chen et al. studied that high-precise and simultaneous monitoring of heavy metals can be monitored with machine learning using plasma spectroscopy \cite{Chen2025MachineLiquids}. Furthermore, ML models integrated with camera-based systems enhance their applicability by facilitating accurate on-site analysis of parameters such as salivary uric acid and pH, while demonstrating adaptability to variations in lighting conditions and image formats \cite{Mutlu2017Smartphone-basedLearning, Liu2024Machine-Learning-basedDetection}.

In this study, we investigated the integration of advanced image processing techniques and ML algorithms to develop an accurate model for detecting $[\text{Ox}]_{\text{tot}}$ based on observable color changes in water samples. We proposed a color-based computer analysis (CBCA) method in which the color change is induced by the oxidant concentration trapped in the KI solution is detected with captured video frames by using a camera system. In this method, a comprehensive image processing methodology was implemented, using seven separate ML models such as LiR, Ridge Regression (RR), RFR, Gradient Boosting Regressor (GBR), and Neural Network (NN) to accurately analyze color binaries in RGB, Lab, and HSV spaces under redox conditions. This approach provided a direct correlation between variations in color density and concentration levels and a reliable dataset for training and validating ML models.

\section{Materials and Method}

\subsection{Experimental Procedure}
\begin{figure}[H]
    \centering
    \includegraphics[width=0.8\linewidth]{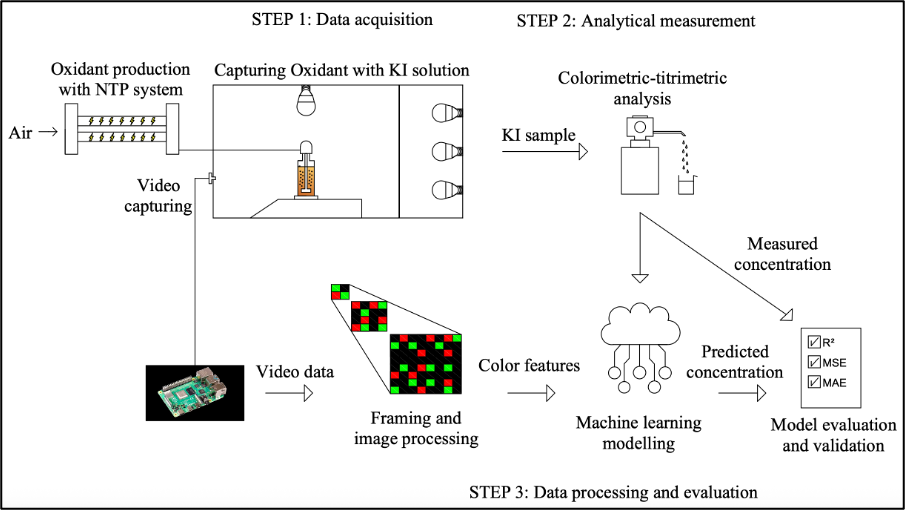}
    \caption{Experimental, analytical and data processing steps}
    \label{fig1}
\end{figure}

The study was carried out in three steps: (1) oxidant production, simultaneously capturing them in a KI solution and recording the video of the KI solution during the oxidation process, (2) measurement of total oxidant concentration with titrimetric KI method at the end of each experimental run, and (3) processing and evaluation of the acquired data in image processing and machine learning steps. Following the experimental and analytical steps (Step~1 and Step~2), the data obtained during the experiments were further processed and evaluated (Step~3) in image processing, machine learning and data evaluation steps (Figure~\ref{fig1}).

\begin{figure}[H]
    \centering
    \includegraphics[width=0.8\linewidth]{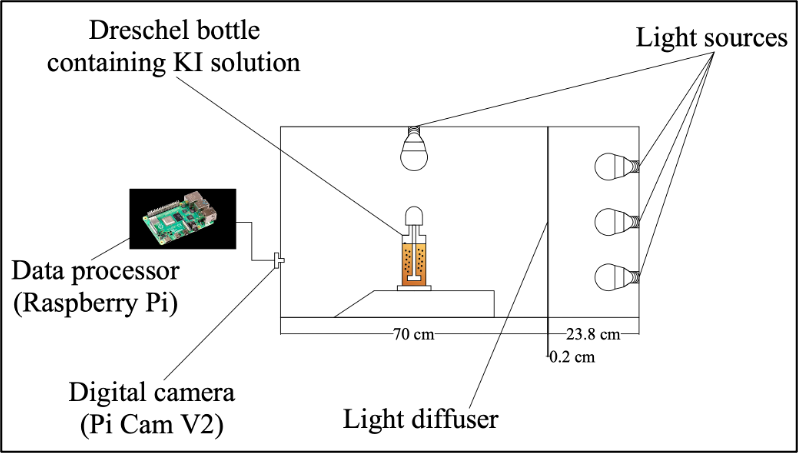}
    \caption{The Genesis visual data acquisition chamber for image capture}
    \label{fig2}
\end{figure}

In the first step, the oxidation and the image data acquisition processes were performed simultaneously. An NTP system was used to generate oxidants (RONS) for obtaining an observable color change in the KI solution. The plasma system was operated at 13~kV of output voltage and 0.8~amp of output current, while 3~L/min of air was fed continuously to the system. At each run, the plasma-treated gas was passed at different durations (1, 6, 9, 14, and 18~minutes) through a 200~mL Drechsel bottle containing 0.12~M KI solution. All oxidation experiments were conducted in triplicate to ensure reproducibility. The Drechsel bottle containing the oxidized KI solution during the experiment had been placed inside a custom-designed visual data acquisition chamber named Genesis, developed in our laboratory (Figure~\ref{fig2}).

The internal dimensions of Genesis are 94~cm (length), 60~cm (width), and 49~cm (height). The chamber is horizontally divided into two functional zones: a 70~cm-long sampling area and a 23.8~cm-long light chamber. These compartments are separated by a polystyrene, 0.2~mm-thick light-diffusion panel to ensure uniform illumination. The light chamber Genesis houses 13 Panasonic 14~W 6500~K bulbs (1521~lumen, 103~mA), while the sampling area is illuminated by four overhead lights to minimize shadows and enhance color consistency. The image/video capturing system contains a Raspberry Pi~4 Model~B with 8~GB RAM and a Raspberry Pi Camera Module~V2 equipped with the Sony IMX219PQH5-C CMOS (Complementary Metal-Oxide Semiconductor) active pixel-type image sensor (diagonal 4.60~mm, Type~1/4.0). The camera is positioned based on the fixed optimal focal distance from the sample. During experiments, KI solution was continuously oxidized by bubbling ozone gas where the solution is inside the Genesis. Rather than capturing still images, videos were recorded in 1280$\times$720 resolution at 30~frames per second (fps) in MP4 format, providing a continuous record of solution behavior. These videos were later decomposed into individual image frames for analysis in the data processing step.

\subsection{Chemical analysis}
In the second step, at the end of each experimental run, 10~mL samples were taken from the Dreschel bottle, and the measurement of $[\text{Ox}]_{\text{tot}}$ was performed according to the KI method given in Standard Methods 2350-E~\cite{ref13}. In chemical analysis, analytical grade sulfuric acid (H$_2$SO$_4$, 96--98\%), sodium thiosulfate (Na$_2$S$_2$O$_3$), and potassium iodide (KI) were used. The 0.12~M KI solution, 0.05~N Na$_2$S$_2$O$_3$, and 2~N H$_2$SO$_4$ solutions were prepared using deionized water (18.2~M$\Omega$, Millipore Milli-Q).

Iodometric titration was conducted at five discrete time points (1, 6, 9, 14, and 18~minutes) for each plasma-treated KI sample to determine total oxidant concentrations ($[\text{Ox}]_{\text{tot}}$). Triplicate titrations were performed to obtain statistically robust measurements. The total oxidant concentration was calculated using Eq.~\ref{eq:ox_tot}:

\begin{equation}
\label{eq:ox_tot}
[\text{Ox}]_{\text{tot}} = \frac{V_{\text{titrant}} \times N \times 24}{V_{\text{sample}}}
\end{equation}

where $V_{\text{titrant}}$ is the volume of sodium thiosulfate used, $N$ is its normality, and $V_{\text{sample}}$ is the volume of the KI solution.

\subsection{Image Processing}
In the third step, the recorded visual data of the KI solution during the oxidation process were subjected to a series of image processing techniques, including segmentation, masking, and color space transformations. These operations enabled the extraction of statistically meaningful color features. At the end of image processing, all the extracted features were used as input variables for machine learning models to predict the total oxidant concentration. For methodological clarity, the pipeline was applied in a fixed sequential order with explicitly parameterized operations.

Initially, the recorded video files were decomposed into image frames using a frame-rate-based extraction method at 30~fps (frames per second). Each video segment, depending on the duration (e.g., 1, 6, 9, 14, 18~minutes), yielded between 1{,}674 and 32{,}277 frames per format. At the end, 257{,}229 raw frame data was obtained in total. For each frame, three image formats (JPEG, PNG, TIFF) were saved to ensure robustness across compression standards.

The image preprocessing pipeline consisted of eight sequential steps applied to the cropped regions of interest (ROIs). The applied order was: (1) raw image loading and BGR-to-RGB conversion, (2) YOLOv8-based RoI segmentation, (3) HSV masking, (4) reflection correction with Telea inpainting, (5) CLAHE enhancement, (6) Gaussian adaptive thresholding, (7) bilateral filtering and near-black pixel removal, and (8) RGB histogram/statistical feature extraction. Initially, raw images were loaded using OpenCV, with conversion from BGR to RGB formats. Next step involves automated segmentation using YOLOv8 \cite{JocherGlenn2023UltralyticsYOLOv8}, a deep-learning model trained on a custom dataset. This segmentation approach enabled the precise identification and extraction of Region of Interest (RoI), minimizing manual intervention and ensuring consistency across all images (details are given in Supplementary Material). After this segmentation step, a series of preprocessing steps were employed in the light of previous studies \cite{Yan2005AMethod, Karim2008LeftAblation, Fatima2020AutomaticImages, Aris2023RobustImages}. Subsequently, HSV masking was applied after converting RGB images into HSV space, where minimum and maximum values for hue, saturation, and value were dynamically calculated. These were used to generate masks that isolate key color features. Bright areas, often resulting from reflections, were identified using HSV thresholds (0, 0, 200) to (180, 255, 255) and corrected using the Telea inpainting algorithm with a radius of 3~pixels.

Contrast enhancement was performed using CLAHE (Contrast Limited Adaptive Histogram Equalization) on the L channel in Lab space, with parameters set to a clip limit of 2.0 and a tile grid size of 8$\times$8, followed by recombination and conversion back to RGB. To suppress noise while preserving edges, bilateral filtering was applied using a 9-pixel diameter, and sigma values of 75 for both color and spatial domains. The main reason to apply CLAHE was to get rid of the bubbles on the data by contrasting them black areas. Subsequently, nearly black pixels were excluded using RGB thresholds ranging from (1, 1, 1) to (255, 255, 255), ensuring that only informative regions contributed to further analysis. Finally, RGB histograms were generated using the remaining masked pixels, with 256 bins per channel and plotted with transparency ($\alpha$ = 0.5) to visualize distributions.

An adaptive thresholding was applied to refine the segmented images and improve robustness under varying illumination conditions. The methods that are used at the processing steps were parameterized to obtain homogeneous data~\cite{Yan2005AMethod,Bhalla2021AnPanels,KaurAnProcessing}. Specifically, Gaussian Adaptive Thresholding was utilized with a block size of 11 and a constant of 2, dynamically adjusting brightness thresholds across different image regions to enhance segmentation accuracy. Bilateral filtering was employed to suppress noise while preserving critical edge details, with a diameter of 9, sigmaColor of 75, and sigmaSpace of 75. The bilateral filtering technique ensured that small variations in density were smoothed without introducing blurring effects. The final feature sets used in model comparison were organized as 9 features, 4 features, 3 features (RGB), and 1 feature (RGB-B), preserving consistency between image processing outputs and machine learning inputs.

\subsection{Data analysis}
Following the image processing pipeline, all cropped and masked images were converted into multiple color spaces, including RGB, HSV, and Lab. This multi-space transformation enabled the extraction of comprehensive colorimetric features representative of the solution’s visible characteristics \cite{Mohan2019IntelligentLight,Carretero-Pena2019EstimationFilters,Moreira2022BenchmarkTomato}. For each 30-frame interval (equivalent to 1~second), the mean values of selected color channels were computed and stored. In the model input table, these second-level records are represented by mean descriptors.

For the purpose of matching the measured experimental $[\text{Ox}]_{\text{tot}}$ values and the extracted color features, the measured experimental $[\text{Ox}]_{\text{tot}}$ values were interpolated using a linear regression model to estimate oxidant concentrations at intermediate time points between experimental runs. A similar methodology has been proposed by Jiang et al., where time-series data augmentation through interpolation was utilized to synthetically generate intermediate data points while preserving the original data structure and variability \cite{Oh2020Time-SeriesInterpolation}. Operationally, this interpolation step was implemented as linear mapping between the start and end concentration values of each run at second resolution. These interpolated values were then matched to the corresponding color features, yielding a final dataset ($n = 464$) suitable for further data analysis and for training and testing the ML models. 

A Pearson correlation matrix and linear regression models were also constructed between the extracted color features and the $[\text{Ox}]_{\text{tot}}$ values to examine dependencies between extracted color features and concentration levels.

\subsection{Model training and evaluation}
Five different ML models were implemented to predict $[\text{Ox}]_{\text{tot}}$ from image-derived color features: Linear Regression (LiR), Ridge Regression (RR), Random Forest Regressor (RFR), Gradient Boosting Regressor (GBR), and Neural Network (NN).

Linear Regression (LiR) was used for establishing a baseline linear model. In the main training pipeline, Ridge Regression (RR) was applied with a fixed regularization strength ($\alpha=1.0$); regularization is used to improve generalizability and reduce overfitting risk in correlated feature spaces \cite{Murphy2012AdaptivePerspective}.

The ensemble-based Random Forest Regressor (RFR) was configured with 100 estimators and a fixed random seed (42) to ensure result reproducibility while maintaining sufficient model depth for reducing variance. Gradient Boosting Regressor (GBR) was applied using default hyperparameters, serving as a reliable baseline for additive tree-based methods due to its robustness and flexibility across small datasets. The Neural Network (NN) was built with two hidden layers of 100 neurons each, chosen as a balanced configuration to allow non-linear pattern learning without excessive parameter growth, and was trained with a maximum of 2000 iterations. Early stopping was activated to prevent overfitting by halting training once validation performance plateaued.

The final dataset ($n = 464$), comprising $[\text{Ox}]_{\text{tot}}$ values and mean color features, was partitioned into training (80\%) and testing (20\%) subsets prior to model input. Input features were normalized using MinMaxScaler to improve model convergence and interpretability.

To ensure consistency and fairness across model comparisons, several common parameters and preprocessing steps were standardized for all machine learning algorithms. Prior to model training, all input features were normalized to a [0, 1] range using the MinMaxScaler, which facilitated faster convergence and improved interpretability. For instance, a large-scale study conducted on 82 datasets demonstrated that improper feature scaling can lead to worse performance, particularly in cross-validation folds, than models trained on unscaled data \cite{deAmorim2022ThePerformance}. The dataset was uniformly partitioned into training (80\%) and testing (20\%) subsets, and five-fold cross-validation (CV) was employed during model evaluation to enhance robustness and mitigate sampling bias. Random state values were fixed to 42 for reproducibility where applicable. All models were trained on predefined feature configurations: full feature set (nine color descriptors), reduced 4-feature set, RGB-only 3-feature set, and single-feature RGB-B set.

First, all the ML models were trained and tested with data from nine color features. Afterwards, comparative training was performed with the reduced feature configurations (4-feature, RGB-only, and RGB-B) to quantify performance changes under dimensionality reduction. The 4-feature subset was selected as the most linearly informative combination after linearity screening, based on univariate feature ranking using $f$-statistics and corresponding $p$-values. The RGB-only subset was additionally evaluated because RGB-based prediction is the most commonly reported configuration in related colorimetric image-analysis studies \cite{GoncalvesDiasDiniz2020ChemometricsassistedSystems, Solmaz2018QuantifyingClassifiers, Jung2022AutomaticSmartphone, Khanal2021Machine-Learning-AssistedDevices}. Finally, the single-channel RGB-B configuration was tested because, within the RGB color space, the B channel showed the strongest linear relationship with $[\text{Ox}]_{\text{tot}}$.

Model performance evaluation was performed after each training-testing session by comparing the coefficient of determination ($R^2$), mean squared error (MSE), and mean absolute error (MAE) of the ML models both on the hold-out test set and under five-fold cross-validation (CV). The five-fold cross validation is a model evaluation technique that splits the training dataset into five equal-sized folds, trains the model on four folds, tests it on the remaining fold, and repeats this process five times to average the performance metrics.

\subsection{Method validation}
An independent set of experiments was performed at the end of the study to validate the accuracy of the proposed CBCA method by comparing the model predictions with the titrimetric $[\text{Ox}]_{\text{tot}}$ measurements obtained via KI method. The validation data was produced using five separate experimental runs, corresponding to the oxidation durations of 2, 5, 8, 12, and 16~min. In each run, 10~mL samples of plasma-treated potassium iodide (KI) solution were collected and analyzed using the KI method carried out by a single trained operator. In parallel, the video recordings of the oxidation processes were processed through image processing pipeline. From the video segments, frames were extracted and processed to gather the color space dataset. This dataset was used as input parameters for model prediction.

\section{Results and Discussion}
\subsection{Determination of total oxidant concentration in KI solution during the experiments}
\begin{figure}[H]
    \centering
    \includegraphics[width=\linewidth,height=0.35\textheight,keepaspectratio]{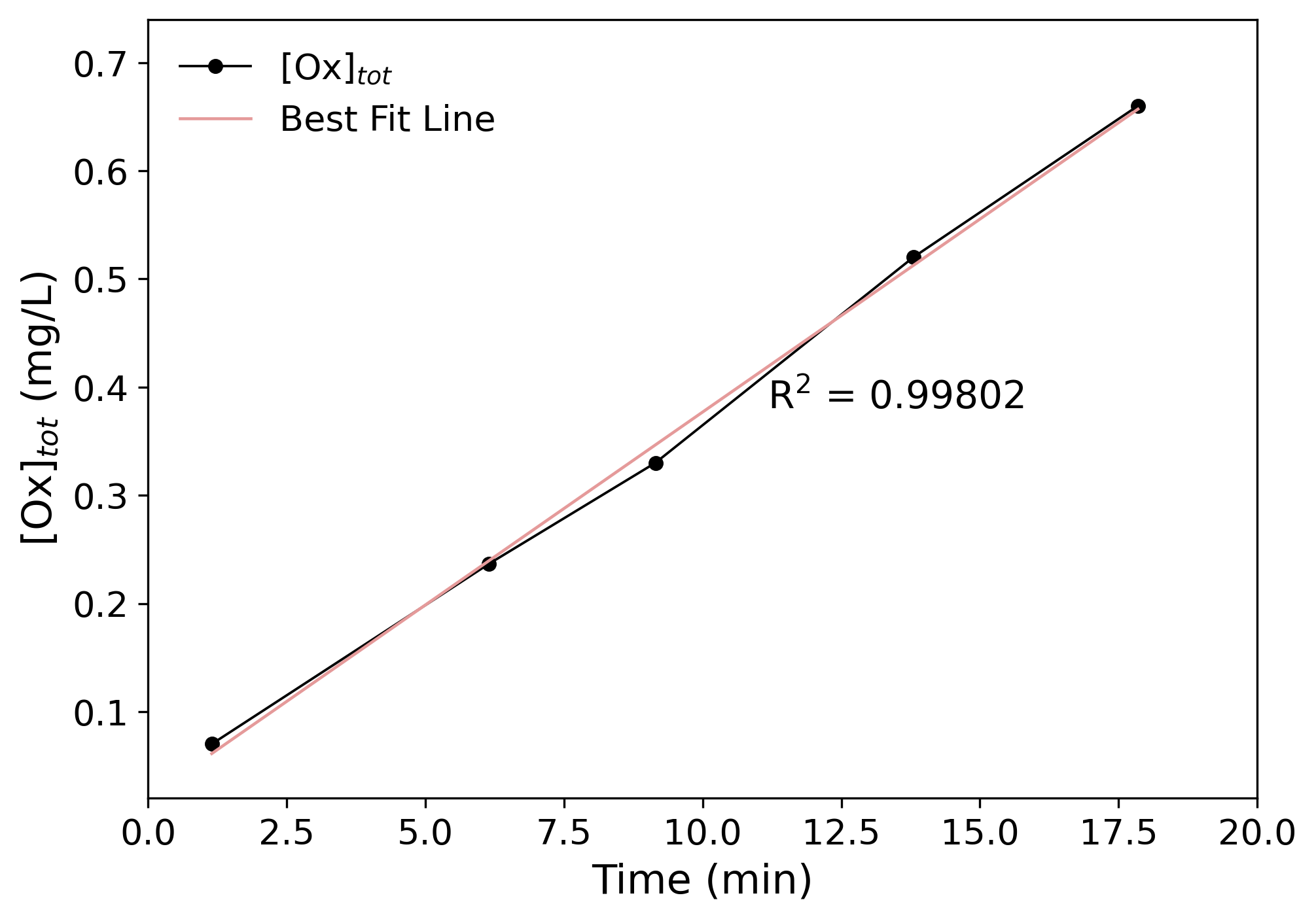}
    \caption{Relationship between the gathered $[\text{Ox}]_{\text{tot}}$ data by titration versus time}
    \label{fig3}
\end{figure}

The total oxidant concentrations $[\text{Ox}]_{\text{tot}}$ values determined through iodometric titration at discrete oxidation intervals---1, 6, 9, 14, and 18~minutes---are presented in Figure \ref{fig3}. For each time point, three replicate measurements were conducted to ensure analytical reliability. The mean oxidant concentrations ranged between 0.33~mg/L and 4.00~mg/L, reflecting the progressive accumulation of oxidants as the plasma exposure time increased. The average analytical precision of the titration protocol was calculated as 0.06 $\pm$ 0.02~mg/L, indicating a high degree of reproducibility across replicates.

\FloatBarrier

A linear regression model was fitted to the time--concentration data, revealing an exceptionally strong correlation between oxidation duration and the total oxidant concentration, with a coefficient of determination of $R^2 = 0.99802$. This outcome not only confirms the temporal predictability of the oxidation process under the applied non-thermal plasma (NTP) conditions but also reinforces the chemical consistency and stability of the iodometric titration method employed.

\subsection{Extracting color features by image processing}
\begin{figure}[H]
    \centering

    % Frame around the whole 2x4 panel block
    {\setlength{\fboxsep}{6pt}% padding
     \setlength{\fboxrule}{0.6pt}% line thickness
     \fbox{%
        \begin{minipage}{0.98\linewidth}
            \centering

            % Row 1 (low concentration)
            \begin{subfigure}[c]{0.49\linewidth}
                \centering
                \includegraphics[width=\linewidth]{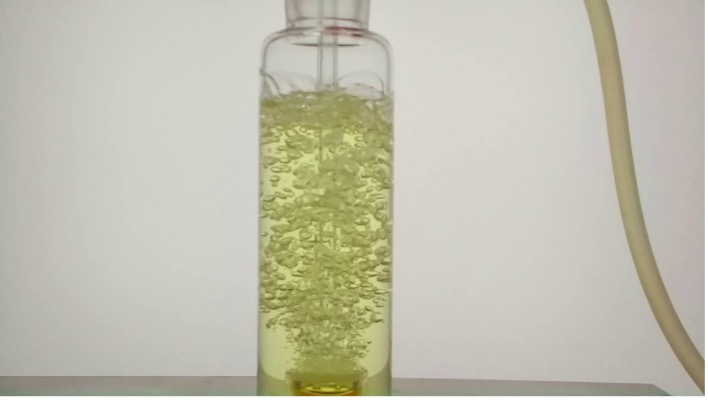}
                \caption*{(a.1)}
            \end{subfigure}\hfill
            \begin{subfigure}[c]{0.16\linewidth}
                \centering
                \includegraphics[width=\linewidth]{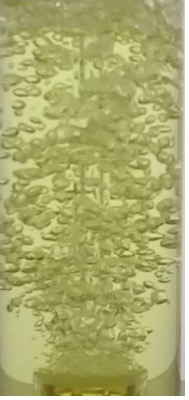}
                \caption*{(b.1)}
            \end{subfigure}\hfill
            \begin{subfigure}[c]{0.16\linewidth}
                \centering
                \includegraphics[width=\linewidth]{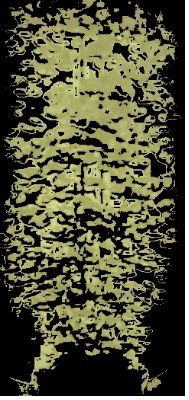}
                \caption*{(c.1)}
            \end{subfigure}\hfill
            \begin{subfigure}[c]{0.16\linewidth}
                \centering
                \includegraphics[width=\linewidth]{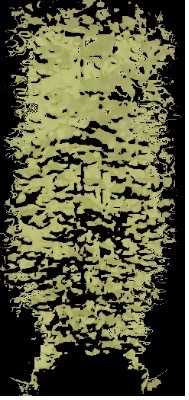}
                \caption*{(d.1)}
            \end{subfigure}

            \vspace{0.8em}

            % Row 2 (high concentration)
            \begin{subfigure}[c]{0.49\linewidth}
                \centering
                \includegraphics[width=\linewidth]{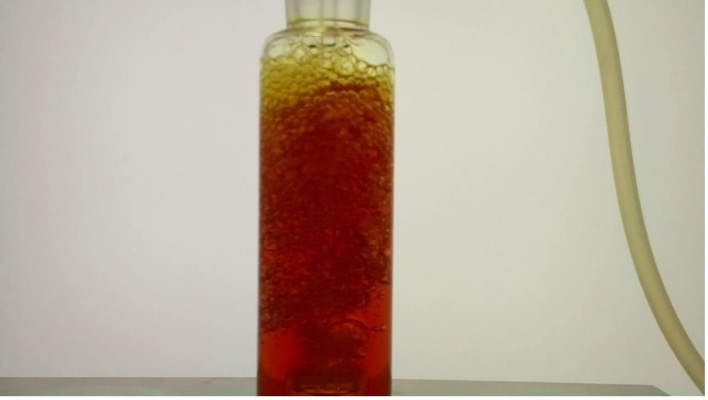}
                \caption*{(a.2)}
            \end{subfigure}\hfill
            \begin{subfigure}[c]{0.16\linewidth}
                \centering
                \includegraphics[width=\linewidth]{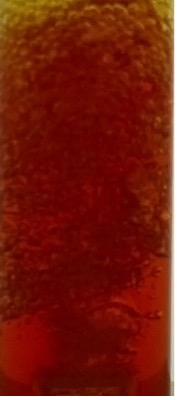}
                \caption*{(b.2)}
            \end{subfigure}\hfill
            \begin{subfigure}[c]{0.16\linewidth}
                \centering
                \includegraphics[width=\linewidth]{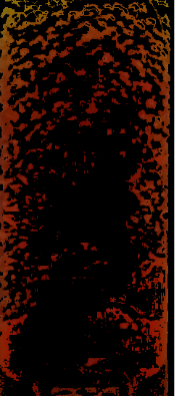}
                \caption*{(c.2)}
            \end{subfigure}\hfill
            \begin{subfigure}[c]{0.16\linewidth}
                \centering
                \includegraphics[width=\linewidth]{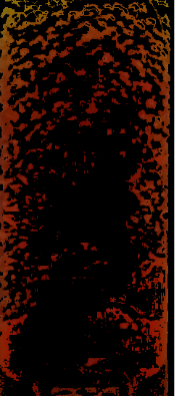}
                \caption*{(d.2)}
            \end{subfigure}
        \end{minipage}%
     }}

    \caption{Image processing pipeline for video frames at two concentration levels ((1) low and (2) high concentration) including (a) the raw images showing the color gradient post-reaction, (b) the cropped region of interest (ROI) based on YOLOv8 detection, (c) the HSV-masked images highlighting the reactive zones, and (d) the final image after artifact removal via Telea inpainting for enhancing analysis accuracy}
    \label{fig4}
\end{figure}

A sequential image-processing pipeline was applied to video frames acquired during continuous oxidation of the KI solution. Figure \ref{fig4} illustrates the initial four stages for two representative concentration regimes: (1) a low concentration level and (2) a high concentration level. In Figure \ref{fig4}a, the raw frame shows the reaction zone together with visible bubble and illumination artifacts. In Figure \ref{fig4}b, the region of interest (RoI) is localized and cropped using the YOLOv8 bounding box, restricting analysis to the reactive zone and reducing background noise. In Figure \ref{fig4}c, HSV masking is applied to isolate informative color regions. In Figure \ref{fig4}d, high-intensity reflection artifacts are mitigated using the Telea inpainting algorithm, with the reflection mask defined in HSV color space as (0,0,200)--(180,255,255) and an inpainting radius of 3 pixels.

\begin{figure}[H]
    \centering

    % Frame around the whole 2x4 panel block
    {\setlength{\fboxsep}{6pt}% padding
     \setlength{\fboxrule}{0.6pt}% line thickness
     \fbox{%
        \begin{minipage}{0.98\linewidth}
            \centering

            % Row 1 (low concentration)
            \begin{subfigure}[c]{0.16\linewidth}
                \centering
                \includegraphics[width=\linewidth]{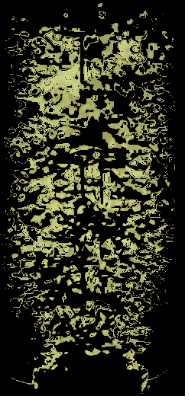}
                \caption*{(a.1)}
            \end{subfigure}\hfill
            \begin{subfigure}[c]{0.16\linewidth}
                \centering
                \includegraphics[width=\linewidth]{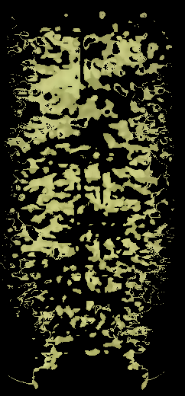}
                \caption*{(b.1)}
            \end{subfigure}\hfill
            \begin{subfigure}[c]{0.16\linewidth}
                \centering
                \includegraphics[width=\linewidth]{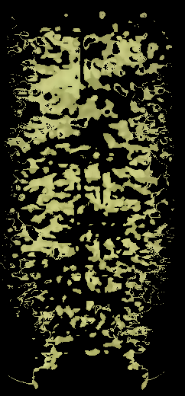}
                \caption*{(c.1)}
            \end{subfigure}\hfill
            \begin{subfigure}[c]{0.49\linewidth}
                \centering
                \includegraphics[width=\linewidth]{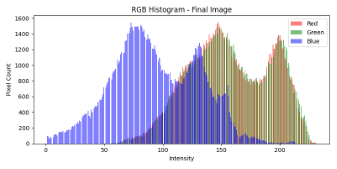}
                \caption*{(d.1)}
            \end{subfigure}

            \vspace{0.8em}

            % Row 2 (high concentration)
            \begin{subfigure}[c]{0.16\linewidth}
                \centering
                \includegraphics[width=\linewidth]{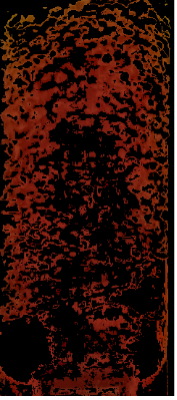}
                \caption*{(a.2)}
            \end{subfigure}\hfill
            \begin{subfigure}[c]{0.16\linewidth}
                \centering
                \includegraphics[width=\linewidth]{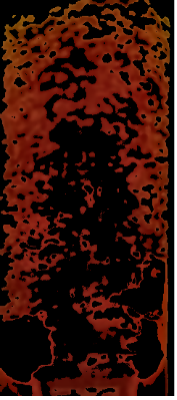}
                \caption*{(b.2)}
            \end{subfigure}\hfill
            \begin{subfigure}[c]{0.16\linewidth}
                \centering
                \includegraphics[width=\linewidth]{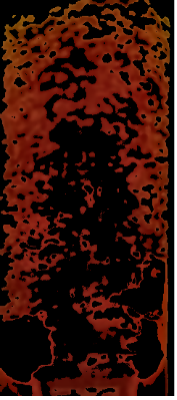}
                \caption*{(c.2)}
            \end{subfigure}\hfill
            \begin{subfigure}[c]{0.49\linewidth}
                \centering
                \includegraphics[width=\linewidth]{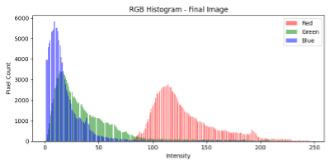}
                \caption*{(d.2)}
            \end{subfigure}
        \end{minipage}%
     }}

    \caption{Image processing steps applied to the reactive ROI for both (1) lower and (2) higher KI concentrations including (a) CLAHE-enhanced images with improved local contrast in the oxidant zone, (b) the images after bilateral filtering, (c) the images after near-black pixel removal, ensuring histogram accuracy, and (d) the final RGB histograms reflect pixel intensity distributions}
    \label{fig5}
\end{figure}

After inpainting, three additional stages were applied to finalize feature extraction (Figure \ref{fig5}). Initially, local contrast was enhanced by applying contrast-limited adaptive histogram equalization (CLAHE) to the L channel in the CIE Lab color space (clip limit = 2.0; tile grid size = $8\times8$), thereby improving the visibility of local oxidant-associated intensity variations (Figure \ref{fig5}a). Second, a bilateral filtering operation (filter diameter = 9, $\sigma_{color} = 75$, $\sigma_{space} = 75$) was applied to attenuate noise while preserving edge structures (Figure \ref{fig5}b). Third, near-black pixels were excluded using the RGB mask range (1,1,1)--(255,255,255) to prevent bias in downstream statistics (Figure \ref{fig5}c). Finally, RGB histograms (256 bins/channel) were generated from the cleaned RoI (Figure \ref{fig5}d), showing concentration-dependent channel distributions.

\subsection{Assessing the distribution of extracted color features}
A violin plot was generated to display the distribution of the mean color feature values from the color spaces including RGB, HSV, and Lab (Figure~6). This visual representation reveals key insights into data spread and variability. Distribution of color spaces were analyzed to assess the underlying distribution characteristics of the extracted features. The most predictive features showed near-normal or bimodal distributions with clear separation, which supports their consistency across experimental conditions. Features with poor regression performance showed wider or skewed distributions, indicating instability or higher noise levels.

\begin{figure}[H]
    \centering
    \includegraphics[width=1\linewidth]{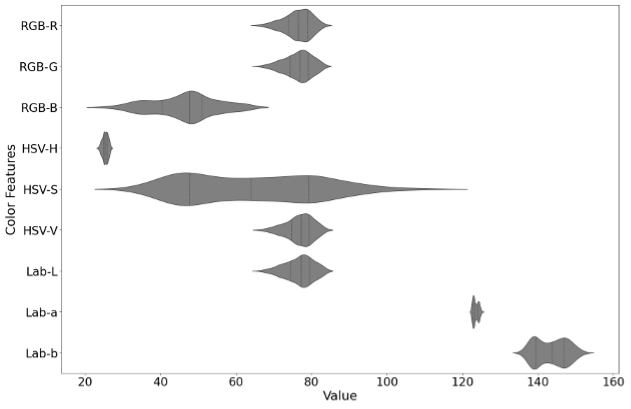}
    \caption{Violin plot of mean color values of the extracted color features across RGB, HSV, and Lab color spaces}
    \label{fig6}
\end{figure}

Within the RGB color space, the red channel (RGB-R) demonstrated a narrow and symmetric distribution centered around higher values (77--79), indicating strong and consistent reddish hues in the oxidized solution. The green channel (RGB-G) showed a moderately broader spread, while the blue channel (RGB-B) exhibited a bimodal distribution, possibly arising from lighting inconsistencies or variations in background reflectance, which may have caused distinct subpopulations to emerge across the image dataset \cite{OnyangoPhysics-basedSoil}.

In the HSV color space, hue (HSV-H) values were tightly clustered around 25--26, reinforcing the chromatic dominance of red-orange tones in the observed samples. In contrast, saturation (HSV-S) presented a significantly broader and asymmetrical distribution, ranging approximately from 40 to above 100. This widespread reflects the varying chromatic intensity due to different stages of the oxidation reaction, spanning from colorless iodide (I$^-$) to intense yellow-brown hues of molecular iodine (I$_2$). The high variability of HSV-S highlights its sensitivity to redox-induced chromatic changes and supports its utility as a discriminative feature in classification models. The distribution of brightness (HSV-V) was more symmetric and quasi-Gaussian in shape, suggesting a consistent illumination profile during image acquisition. However, both saturation (HSV-S) and brightness (HSV-V) in the HSV color space are known to behave non-linearly under varied lighting or reflective surface conditions \cite{Joh2025AnAssays}, which supports the skewness observed in the saturation profile.

In the Lab color space, lightness (Lab-L) showed a narrow and symmetric distribution, implying uniform sample brightness. The Lab-a channel (green--red axis) was concentrated around moderately high values (123--127), reflecting the red-shifted nature of the oxidized solution. The most striking variability occurred in the Lab-b channel (blue--yellow axis), which not only had high magnitude values but also a broad, possibly bimodal distribution. This aligns with the dominance of yellow tones in the oxidized samples and underscores the Lab-b channel\textquoteright s responsiveness to chromatic shifts associated with iodine redox transitions. Such multimodal distribution likely indicates the existence of distinct optical states or sample batches \cite{Ojeda2023UnderstandingImages}.

Collectively, among all analyzed features, RGB-B, HSV-S and Lab-b stood out with the broadest and most variable distributions. These three features were highly responsive to oxidative conditions and exhibit strong potential as primary indicators for both reaction tracking and machine learning-based classification models. Conversely, HSV-H and Lab-a displayed the least variability, which may limit their predictive usefulness due to their narrow and stable distributions across experiments \cite{Rengasamy2022FeatureApproach}; however, low-variance features can still provide meaningful predictive power when they maintain a strong and consistent relationship with the target variable, as their limited fluctuation does not preclude informativeness within that range \cite{Chidambaram2025ForAugmentation}.

\subsection{Analyzing the correlation between the color features and the oxidant concentrations}
\begin{figure}
    \centering
    \includegraphics[width=1\linewidth]{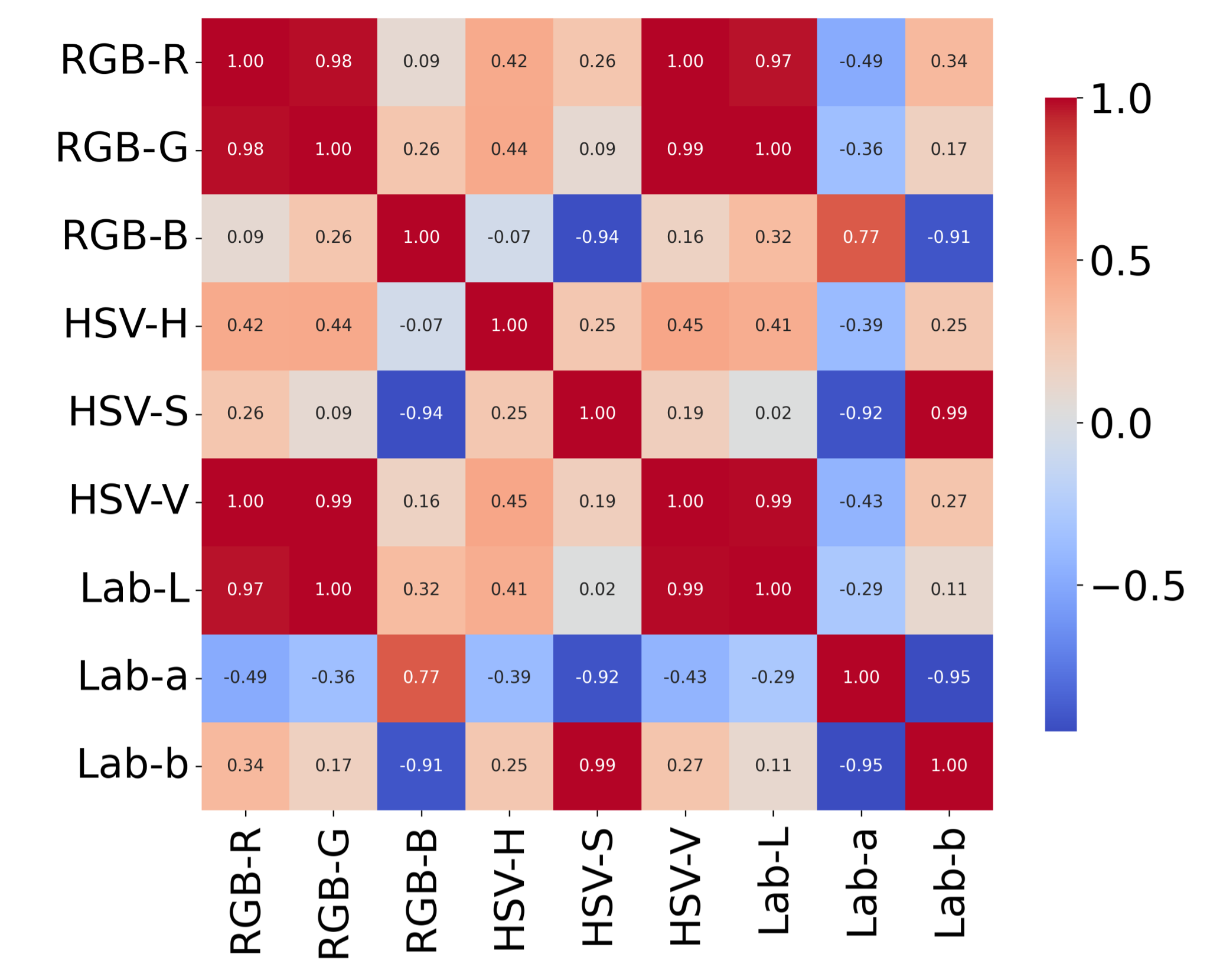}
    \caption{Pearson correlation matrix showing relationships among extracted color features and $[\text{Ox}]_{\text{tot}}$}
    \label{fig7}
\end{figure}

A Pearson correlation matrix was constructed to understand the statistical dependencies between the extracted image-based color features and the experimentally determined oxidant concentrations (Figure \ref{fig7}). The heatmap reveals how strongly each color channel correlates with the $[\text{Ox}]_{\text{tot}}$ values and with each other, offering crucial insights for model feature selection and interpretability.

Notably, the saturation channel in HSV space (HSV-S) exhibited the highest positive correlation with $[\text{Ox}]_{\text{tot}}$ ($r = 0.96$), followed closely by the Lab blue--yellow axis (Lab-b, $r = 0.97$). This indicates that as the concentration of oxidants increases, both saturation and the blue--yellow component in Lab consistently intensify, supporting their utility in oxidative color tracking. Conversely, green (RGB-G) and blue (RGB-B) channels from RGB space were negatively correlated with the $[\text{Ox}]_{\text{tot}}$ ($r = -0.93$), suggesting that oxidant-induced color transitions are associated with the suppression of cooler hues. Notably, this aligns with results by Tarim \& Tekin (2024), who demonstrated that HSV-S often exhibits the strongest correlation with analyte concentration in colorimetric assays, outperforming RGB components---further validating our observed $r = 0.96$ relationship between HSV-S and the $[\text{Ox}]_{\text{tot}}$ \cite{Tarim2024ColorimetricAnalysis}.

The red channel (RGB-R) showed a moderate positive correlation ($r = 0.33$), while the hue (HSV-H) and brightness (HSV-V) channels displayed weak correlations, indicating limited predictive relevance under the studied conditions. Lab-L and Lab-a also had marginal relationships with $[\text{Ox}]_{\text{tot}}$ levels. Additionally, strong multicollinearity was observed between several features---such as RGB-R and HSV-V ($r = 1.00$), and HSV-S and Lab-b ($r = 0.99$)---which informs dimensionality reduction strategies in model training.

These correlations support the hypothesis that selective channel extraction, particularly from HSV and Lab domains, can enhance model sensitivity and reduce noise from irrelevant features, ultimately improving the performance of machine learning models for colorimetric oxidant determination.

To further quantify the relationship between each color feature and the $[\text{Ox}]_{\text{tot}}$, a series of linear regression analyses were performed as presented in Figure~8. The saturation channel (HSV-S) demonstrated the strongest linear relationship ($R^2 = 0.978$), followed by the Lab blue--yellow component (Lab-b, $R^2 = 0.963$) and the Lab a-channel (Lab-a, $R^2 = 0.794$). The blue channel from RGB (RGB-B) also showed a strong inverse relationship with oxidant levels ($R^2 = 0.986$), consistent with earlier correlation results. In contrast, other features---such as brightness (HSV-V, $R^2 = 0.282$), hue (HSV-H, $R^2 = 0.059$), green channel (RGB-G, $R^2 = 0.027$), and Lab lightness (Lab-L, $R^2 = 0.000$)---exhibited low predictive power in the regression model, indicating weak or negligible linear associations with $[\text{Ox}]_{\text{tot}}$. These results confirm that not all channels contribute equally to predictive modeling.

\clearpage
\begin{landscape}
\thispagestyle{empty}
\begin{figure}[p]
    \centering
    \includegraphics[width=0.95\linewidth,height=0.95\textheight,keepaspectratio]{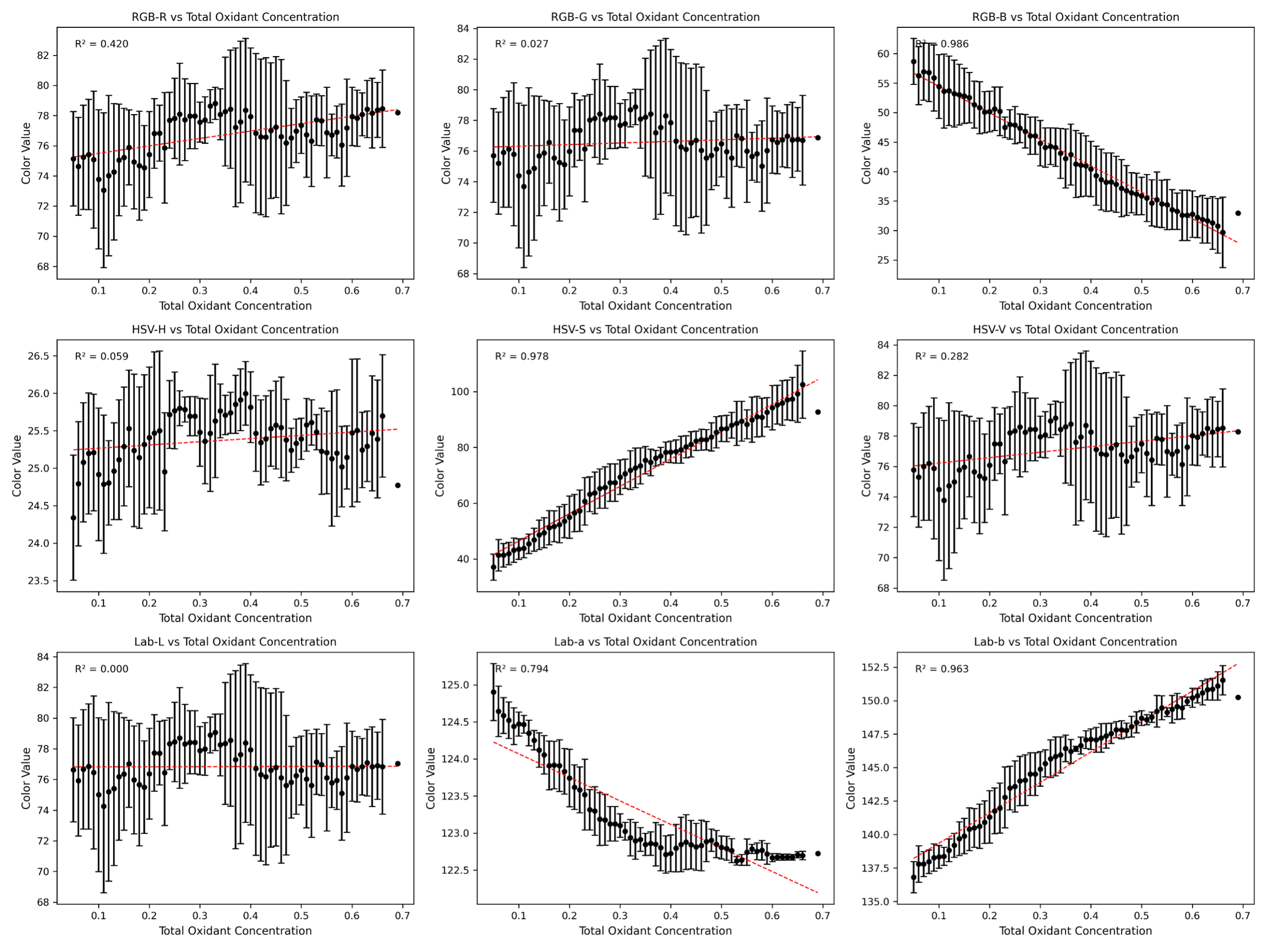}
    \caption{Linear regression plots including the relationship between color features and $[\text{Ox}]_{\text{tot}}$}
    \label{fig8}
\end{figure}
\end{landscape}
\clearpage

These analyses have direct implications for machine learning (ML) model design. Features with high $R^2$ values and well-defined distributions are likely to enhance the generalization capability and predictive strength of ML models. Conversely, including poorly correlated or noisy features may introduce overfitting risks, increase training complexity, or reduce model interpretability. Notably, although the Lab-a feature exhibited a relatively strong linear regression performance with an $R^2 > 0.80$, as shown in Figure~8, its trend more closely resembles a logarithmic relationship rather than a truly linear relationship. Therefore, feature selection informed by both statistical correlation and regression performance serves as a critical preprocessing step to improve the efficiency, robustness, and accuracy of ML-based oxidant concentration determination systems.

In addition, machine learning models were independently trained both with and without the inclusion of the Lab-a feature to assess its impact on model performance (data not shown). Comparative results demonstrated that incorporating Lab-a consistently led to higher $R^2$ values across multiple regression models, indicating an improvement in predictive accuracy. To further evaluate the influence of this feature, post hoc explainability analyses were performed using SHAP (SHapley Additive exPlanations) and LIME (Local Interpretable Model-Agnostic Explanations) (see Supplementary Material). These methods confirmed that Lab-a contributes significantly to the model\textquoteright s output, reinforcing the decision to retain this channel in the final feature set for enhanced model interpretability and performance.

\subsection{Performance evaluation of machine learning models}

\subsubsection{Evaluation of models trained with 9-features}
Model performance was evaluated under four independent dataset configurations in order to systematically compare the influence of feature dimensionality and color-space restriction on predictive behavior. These configurations consisted of: (i) a nine-feature dataset including all RGB, HSV, and Lab channels; (ii) a four-feature dataset obtained after statistical feature selection (RGB-B, HSV-S, Lab-a, Lab-b); (iii) a three-feature dataset composed exclusively of RGB-R, RGB-G, and RGB-B channels; and (iv) a one-feature dataset using only RGB-B.

All four datasets were trained and tested separately using identical preprocessing steps (MinMaxScaler normalization), identical train--test partitioning (80\% training, 20\% testing), and identical five-fold cross-validation procedures, ensuring methodological consistency across experiments.

A comprehensive performance evaluation of five machine learning models---including Linear Regression (LiR), Ridge Regression (RR), Random Forest Regression (RFR), Gradient Boosting Regressor (GBR), and Neural Networks---was conducted using the refined feature set. Table~1 presents the evaluation metrics obtained under the nine-feature configuration (RGB, HSV, Lab), which represents the most information-rich dataset among the four experimental scenarios and serves as the reference baseline for subsequent comparisons.

\begin{table}[H]
\centering
\caption{Evaluation metrics ($R^2$, MSE, and MAE) of ML models (LiR, RR, RFR, GBR, and NN) assessing their performance on color-based data.}
\label{tab:ml_metrics}
\small
\setlength{\tabcolsep}{5pt}
% Force the table to stay centered even if it needs slight scaling
\resizebox{\linewidth}{!}{%
\begin{tabular}{lccccccc}
\hline
Model & $R^2$ (Train) & $R^2$ (Test) & MSE & MAE & CV $R^2$ Mean & CV $R^2$ Std & Training time (ms)\\
\hline
LiR & 0.9843 & 0.9904 & 0.0003 & 0.0137 & 0.9823 & 0.0043 & 0.3889\\
RR & 0.9604 & 0.9784 & 0.0007 & 0.0228 & 0.9552 & 0.0067 & 0.3541\\
RFR & 0.9968 & 0.9853 & 0.0005 & 0.0175 & 0.9812 & 0.0075 & 124.3269\\
GBR & 0.9988 & 0.9913 & 0.0003 & 0.0145 & 0.9820 & 0.0067 & 26.1939\\
NN & 0.9468 & 0.9675 & 0.0010 & 0.0269 & 0.9458 & 0.0387 & 92.6981\\
\hline
\end{tabular}%
}
\end{table}

These performance metrics provide complementary perspectives on model behavior. $R^2$ quantifies how well each model explains the variance in the target variable and is a widely used indicator of overall fit. Mean squared error (MSE), which penalizes larger errors more heavily due to squaring, is particularly sensitive to outliers and provides insight into model precision in terms of variance. Mean absolute error (MAE), on the other hand, offers a more intuitive interpretation by representing the average magnitude of error across predictions, making it useful for assessing typical prediction accuracy. Cross-validated $R^2$ scores and their standard deviations assess the consistency and generalizability of the models across multiple data splits. Finally, training time provides a practical assessment of model scalability and suitability for time-sensitive applications.

Among the models, the Gradient Boosting Regressor achieved the highest predictive performance with an $R^2$ of 0.9912, the lowest MSE of 0.000274, and a relatively low MAE of 0.0146. This performance trend is consistent with results by Borisov et~al. (2023), who demonstrated that Gradient Boosting methods consistently outperform neural networks in structured datasets, especially under limited data conditions \cite{Borisov2022DeepSurvey}. The low MSE value indicates minimal squared deviation from actual $[\text{Ox}]_{\text{tot}}$ values, reflecting excellent pointwise prediction consistency, while the low MAE signifies strong average predictive precision across the dataset. Linear Regression also showed strong generalization ($R^2 = 0.9904$), supported by an impressively low MSE of 0.000299 and MAE of 0.0137---indicating that despite its simplicity, it delivered both accurate and consistent predictions. Additionally, it exhibited the fastest training time (0.0019~s), making it highly efficient for real-time or large-scale applications. This further highlights the efficiency and robustness of linear models in scenarios with low-noise, high-quality input features, especially when model interpretability and computational cost are key constraints (see Supplementary Material).

\subsubsection{Evaluation of models trained with 4-features}
To enhance model interpretability and prevent overfitting, a statistical feature selection process was implemented using univariate linear regression tests based on comparing F-scores and $p$-values. The resulting F-scores and associated $p$-values quantify each feature\textquoteright s individual explanatory power regarding the target variable ($[\text{Ox}]_{\text{tot}}$). Figure~9 visualizes these F-scores and the $p$-values for all extracted color features.

\begin{figure}[H]
    \centering
    \includegraphics[width=0.9\linewidth]{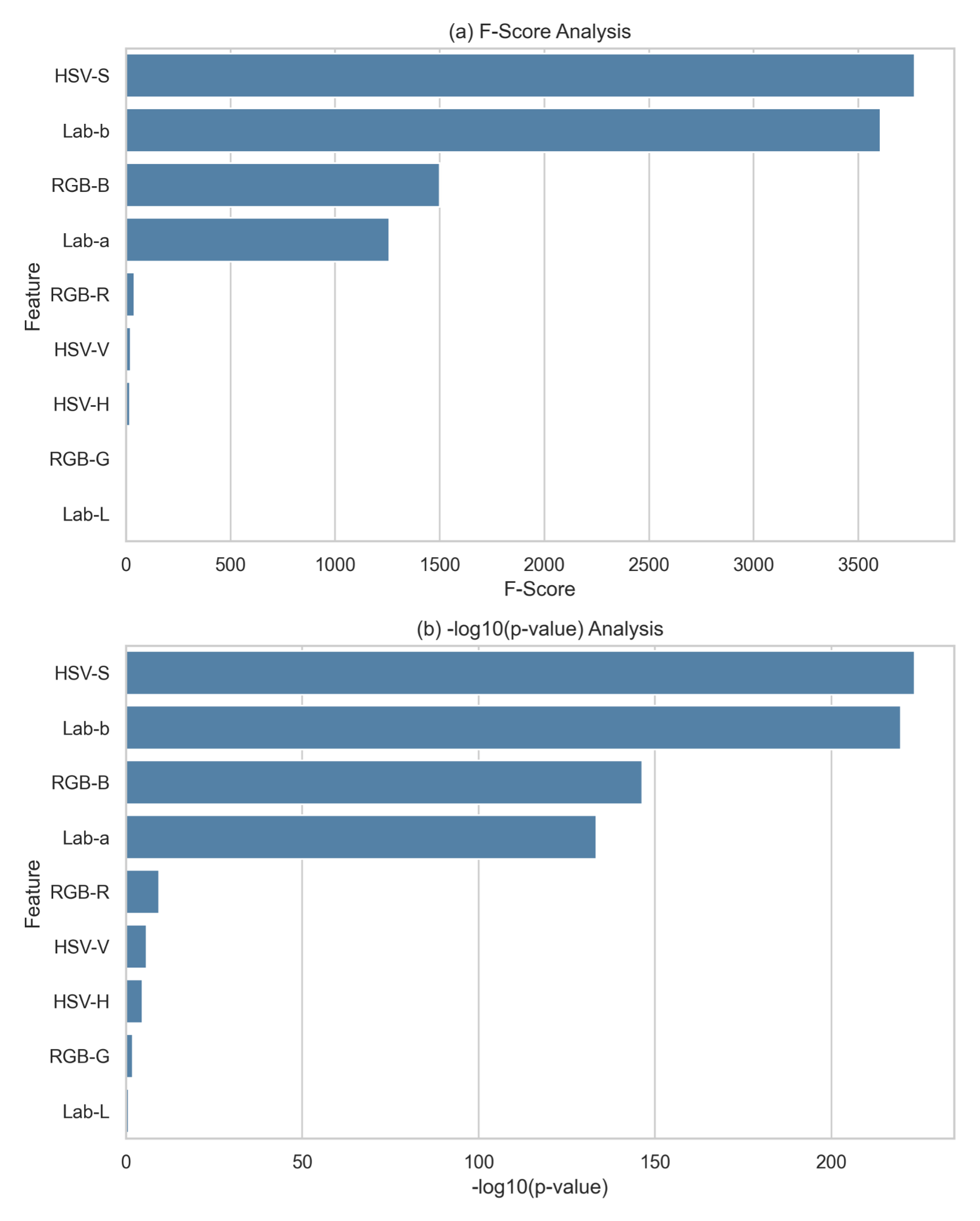}
    \caption{Bar chart of F-scores and p-values from univariate feature selection, ranking color features by their individual predictive power for oxidant concentration}
    \label{fig9}
\end{figure}

The saturation component in HSV space (HSV-S) and the blue--yellow channel in Lab space (Lab-b) emerged as the most significant predictors, with F-scores of 3769.21 and 3605.92, respectively, and extremely low $p$-values ($\approx 10^{-220}$), confirming their statistical importance. These features not only showed strong linear correlation and high regression $R^2$ but were also validated as statistically robust predictors.

Other features like RGB-B ($F = 1499.53$) and Lab-a ($F = 1259.65$) also demonstrated substantial explanatory power, suggesting they capture complementary colorimetric trends during the redox transition. In contrast, features such as RGB-G ($F = 6.71$, $p = 0.0099$), Lab-L ($F = 1.88$, $p = 0.171$), and HSV-H ($F = 18.55$, $p = 0.00002$) were found to be statistically weak and less informative in predicting oxidant levels.

A comparative evaluation of machine learning models trained on the full set of color-space features (RGB, HSV, Lab) versus a statistically selected feature set (RGB-B, HSV-S, Lab-b, and Lab-a) revealed notable differences in performance dynamics, as shown in Figure~10. Linear and Ridge Regression models exhibited the most pronounced declines in accuracy (0.5\% and 2.81\%, respectively), indicating a higher dependence on the full feature space to capture linear relationships and mitigate multicollinearity effects. This sensitivity aligns with the known impact of multicollinearity on ordinary and Ridge Regression models, where removing correlated features can degrade linear fit quality despite improving coefficient stability \cite{Saputro2025PerformanceMulticollinearity}. While Linear Regression\textquoteright s test $R^2$ slightly decreased after feature reduction (from 0.9904 to 0.9854), the lower variance in training and validation errors suggests improved generalization, likely due to the elimination of multicollinearity and redundant inputs. Conversely, tree-based ensemble models such as Random Forest and Gradient Boosting showed minimal performance loss (0.75\%), reflecting their robustness to feature reduction due to their inherent capability to prioritize informative splits. Across models, feature selection consistently yielded reductions in training time, while maintaining acceptable generalization error and cross-validation stability. Similar improvements in ensemble models---maintaining predictive power while reducing training time and eliminating redundant features---have been demonstrated through Gradient Boosted Feature Selection, which achieves scalable, efficient selection of informative features in high-dimensional datasets \cite{Xu2019GradientSelection}. These results suggest that a carefully selected subset of statistically significant features can preserve, or even enhance, predictive performance in complex models, while also improving computational efficiency---making it an effective strategy for streamlined yet accurate oxidant concentration prediction systems.

Table~2 presents the performance metrics obtained under the four-feature configuration (RGB-B, HSV-S, Lab-a, Lab-b), enabling direct comparison with the nine-feature baseline while preserving identical modeling and validation conditions. These features were chosen based on their high F-scores and $p$-values, strong correlation coefficients, and consistent regression performance.

\begin{table}[H]
\centering
\caption{Evaluation metrics ($R^2$, MSE, and MAE) of ML models (LiR, RR, RFR, GBR, and NN) assessing their performance on color-based data after feature selection.}
\label{tab:ml_metrics_fs}
\small
\setlength{\tabcolsep}{5pt}
\resizebox{\linewidth}{!}{%
\begin{tabular}{lccccccc}
\hline
Model & $R^2$ (Train) & $R^2$ (Test) & MSE & MAE & CV $R^2$ Mean & CV $R^2$ Std & Training time (ms)\\
\hline
LiR & 0.9697 & 0.9854 & 0.0005 & 0.0186 & 0.9678 & 0.0048 & 9.6860\\
RR & 0.9242 & 0.9596 & 0.0013 & 0.0279 & 0.9203 & 0.0153 & 5.8961\\
RFR & 0.9963 & 0.9838 & 0.0005 & 0.0188 & 0.9771 & 0.0081 & 60.2748\\
GBR & 0.9983 & 0.9839 & 0.0005 & 0.0181 & 0.9786 & 0.0071 & 53.3187\\
NN & 0.8942 & 0.9471 & 0.0016 & 0.0338 & 0.9201 & 0.0141 & 151.0370\\
\hline
\end{tabular}%
}
\end{table}

Tree-based methods like Random Forest demonstrated slightly lower $R^2$ (0.9851), accompanied by an MSE of 0.000463 and an MAE of 0.0176, which are slightly elevated compared to boosting or linear models. Nonetheless, these values remain acceptable for practical use and indicate reliable performance with some added robustness due to the ensemble averaging. Ridge Regression yielded robust but slightly less accurate results ($R^2 = 0.9877$), with an MSE of 0.000381 and MAE of 0.0159. These metrics suggest that selecting L2 regularization instead of L1 helped manage overfitting while maintaining a reasonable error margin, though the added complexity did not significantly outperform the baseline Linear Regression. In models incorporating normalization techniques such as Batch Normalization, L2 regularization primarily influences the scale of model weights rather than acting as a conventional means of controlling overfitting. This effect contrasts with L1 regularization, which tends to eliminate correlated features and may reduce predictive capacity in structured datasets. As demonstrated by van~Laarhoven (2017), L2 regularization alters the effective learning rate by modulating weight magnitudes, supporting its use in contexts where feature retention and training stability are prioritized \cite{vanLaarhoven2017L2Normalization}.

The neural network model exhibited lower predictive accuracy in this context. The fully connected Neural Network recorded an $R^2$ of 0.9675, an MSE of 0.001008, and an MAE of 0.0269---indicating a noticeable decline in both pointwise and average predictive accuracy. These results reflect increased sensitivity to data noise and suggest overfitting or undertraining may be contributing factors.

Cross-validation scores (CV $R^2$) confirmed model robustness, with Gradient Boosting and Linear Regression maintaining low standard deviation across folds, indicating stability. In contrast, Neural Networks showed the highest CV $R^2$ standard deviation (0.0387), reflecting overfitting due to limited training data or suboptimal hyperparameter tuning, both of which are common challenges in training deep learning models on structured datasets. This is a well-documented behavior in deep learning applied to tabular data, where neural architectures tend to overfit without significant regularization or tuning, particularly when sample size is small or feature noise is high \cite{Shwartz-Ziv2021TabularNeed}.

Overall, the results suggest that tree-based ensemble methods (Random Forest and Gradient Boosting) and linear approaches (Linear and Ridge Regression) offer the most effective balance between predictive accuracy, computational efficiency, and model interpretability. This pattern is consistent with the bias--variance trade-off principle, whereby simpler or regularized models---such as Linear and Ridge Regression---tend to generalize more reliably and yield stable validation metrics, especially when data quantity is limited. While neural networks demonstrated some predictive capability, their performance was notably more sensitive to training variability, reflecting known limitations in applying deep learning to structured, low-noise datasets without substantial tuning and data scaling. The results also emphasize the importance of using statistically significant features with strong correlation to the target variable, as these enhance generalization and reduce the risk of overfitting. In contrast, the inclusion of noisy or weakly correlated inputs may increase model complexity, compromise interpretability, and degrade predictive reliability. Thus, feature selection guided by correlation metrics and regression-based importance scores proves essential to optimizing the robustness and efficiency of ML-driven oxidant concentration prediction workflows.

\subsubsection{Evaluation of models trained with 3-features}

To further investigate the impact of restricting the feature space to a single color space, models were trained using only the three RGB channels (R, G, and B). This configuration reflects one of the most widely adopted color representations in computer vision applications and consumer-grade imaging systems, and therefore serves as a practically relevant constraint scenario. Unlike the nine-feature (RGB+HSV+Lab) configuration, this setup eliminates perceptual and chromaticity-normalized representations (HSV and Lab), enabling an assessment of how much predictive power is retained when operating solely within the native acquisition space of the camera sensor.

Table~\ref{tab:ml_metrics_rgb3} presents the evaluation metrics obtained under the three-feature configuration. Compared with the nine-feature baseline, a modest reduction in predictive accuracy is observed for linear models. Linear Regression achieved a test $R^2$ of 0.9817, with an MSE of 0.0006 and an MAE of 0.0195. While this represents a slight decline relative to the full feature configuration ($R^2 = 0.9904$), the performance remains robust, indicating that a substantial proportion of the variance in $[\text{Ox}]_{\text{tot}}$ is already captured within the RGB space. The relatively low CV standard deviation (0.0069) further confirms stable generalization across folds.

\begin{table}[H]
\centering
\caption{Evaluation metrics ($R^2$, MSE, and MAE) of ML models (LiR, RR, RFR, GBR, and NN) for 3-feature RGB data.}
\label{tab:ml_metrics_rgb3}
\small
\setlength{\tabcolsep}{5pt}
\resizebox{\linewidth}{!}{%
\begin{tabular}{lccccccc}
\hline
Model & $R^2$ (Train) & $R^2$ (Test) & MSE & MAE & CV $R^2$ Mean & CV $R^2$ Std & Training time (ms)\\
\hline
LiR & 0.9570 & 0.9817 & 0.0006 & 0.0195 & 0.9528 & 0.0069 & 0.3781\\
RR  & 0.9317 & 0.9711 & 0.0009 & 0.0195 & 0.9244 & 0.0125 & 0.3440\\
RFR & 0.9956 & 0.9874 & 0.0004 & 0.0150 & 0.9723 & 0.0065 & 80.2701\\
GBR & 0.9977 & 0.9911 & 0.0003 & 0.0124 & 0.9769 & 0.0054 & 102.6440\\
NN  & 0.9358 & 0.9752 & 0.0008 & 0.0223 & 0.3159 & 0.3087 & 335.9032\\
\hline
\end{tabular}%
}
\end{table}

Ridge Regression exhibited a more pronounced reduction in performance ($R^2 = 0.9711$), suggesting that under reduced feature dimensionality, L2 regularization may introduce additional bias without compensatory gains in variance reduction. This behavior is consistent with bias--variance trade-off principles: when informative nonlinear interactions are not explicitly represented, additional regularization can degrade model fit.

In contrast, tree-based ensemble models maintained high predictive performance. Random Forest achieved a test $R^2$ of 0.9874 (MSE = 0.0004), while Gradient Boosting reached $R^2 = 0.9911$ with the lowest MAE (0.0124) among all models in this configuration. Notably, the performance of Gradient Boosting under RGB-only input is nearly indistinguishable from the nine-feature scenario, implying that nonlinear interactions among R, G, and B channels sufficiently encode the dominant chromatic variation associated with oxidant concentration. This finding indicates that the added HSV and Lab transformations contribute limited incremental predictive information when ensemble learners are employed.

The Neural Network model, although achieving a reasonable test $R^2$ of 0.9752, demonstrated substantial instability across folds (CV $R^2$ mean = 0.3159; std = 0.3087). This sharp discrepancy between hold-out test performance and cross-validation behavior suggests sensitivity to data partitioning and potential overfitting under limited sample size. Such instability is well documented in deep learning applied to structured, low-dimensional tabular datasets, where neural architectures often require extensive regularization and hyperparameter tuning to achieve reliable generalization.

From a computational perspective, linear models retained their negligible training time ($< 1$~ms), whereas ensemble methods required substantially higher computational cost (80--102~ms). Nonetheless, given the minimal difference in predictive accuracy between three-feature and nine-feature configurations for Gradient Boosting, the RGB-only representation emerges as a computationally efficient and practically scalable alternative for real-time oxidant monitoring systems.

Overall, the three-feature results indicate that the primary predictive signal resides within the RGB space itself. While multi-space representations (HSV and Lab) marginally enhance linear model performance, ensemble methods effectively exploit nonlinear interactions within RGB channels, achieving near-baseline accuracy without requiring expanded color-space transformations. This supports the methodological choice of evaluating RGB-restricted models as an intermediate reduction step between full feature integration and single-channel minimalism.

\begin{table}[H]
\centering
\caption{Evaluation metrics ($R^2$, MSE, and MAE) of ML models (LiR, RR, RFR, GBR, and NN) for 3-feature RGB data.}
\label{tab:ml_metrics_rgb3}
\small
\setlength{\tabcolsep}{5pt}
\resizebox{\linewidth}{!}{%
\begin{tabular}{lccccccc}
\hline
Model & $R^2$ (Train) & $R^2$ (Test) & MSE & MAE & CV $R^2$ Mean & CV $R^2$ Std & Training time (ms)\\
\hline
LiR & 0.9570 & 0.9817 & 0.0006 & 0.0195 & 0.9528 & 0.0069 & 0.3781\\
RR  & 0.9317 & 0.9711 & 0.0009 & 0.0195 & 0.9244 & 0.0125 & 0.3440\\
RFR & 0.9956 & 0.9874 & 0.0004 & 0.0150 & 0.9723 & 0.0065 & 80.2701\\
GBR & 0.9977 & 0.9911 & 0.0003 & 0.0124 & 0.9769 & 0.0054 & 102.6440\\
NN  & 0.9358 & 0.9752 & 0.0008 & 0.0223 & 0.3159 & 0.3087 & 335.9032\\
\hline
\end{tabular}%
}
\end{table}

\subsubsection{Evaluation of models trained with 1-feature}

To establish the minimal information scenario, models were trained using only the RGB-B channel, which previously exhibited the strongest individual linear association with $[\text{Ox}]_{\text{tot}}$ in the univariate analysis. This configuration represents an intentional reduction to a single explanatory variable and therefore functions as a lower-bound performance benchmark, analogous to a classical single-wavelength calibration approach in colorimetric analysis.

As shown in Table~\ref{tab:ml_metrics_rgbb1}, restricting the feature space to a single channel leads to a clear reduction in predictive performance across all models. Linear Regression achieved a test $R^2$ of 0.8820 (MSE = 0.0037; MAE = 0.0512), indicating that although RGB-B retains substantial explanatory power, it does not capture the full variability observed in multi-channel configurations. The corresponding cross-validation mean ($R^2 = 0.8418$; std = 0.0294) suggests moderate generalization but increased sensitivity to data partitioning relative to higher-dimensional setups.

\begin{table}[H]
\centering
\caption{Evaluation metrics ($R^2$, MSE, and MAE) of ML models (LiR, RR, RFR, GBR, and NN) for 1-feature RGB-B data.}
\label{tab:ml_metrics_rgbb1}
\small
\setlength{\tabcolsep}{5pt}
\resizebox{\linewidth}{!}{%
\begin{tabular}{lccccccc}
\hline
Model & $R^2$ (Train) & $R^2$ (Test) & MSE & MAE & CV $R^2$ Mean & CV $R^2$ Std & Training time (ms)\\
\hline
LiR & 0.8491 & 0.8820 & 0.0037 & 0.0512 & 0.8418 & 0.0294 & 2.5153\\
RR  & 0.8388 & 0.8962 & 0.0032 & 0.0430 & 0.8278 & 0.0265 & 0.3932\\
RFR & 0.9858 & 0.9287 & 0.0022 & 0.0304 & 0.9212 & 0.0235 & 108.4368\\
GBR & 0.9823 & 0.9096 & 0.0028 & 0.0322 & 0.9196 & 0.0200 & 16.1483\\
NN  & 0.7779 & 0.8540 & 0.0045 & 0.0568 & 0.3775 & 0.5592 & 179.2510\\
\hline
\end{tabular}%
}
\end{table}

Notably, Ridge Regression slightly outperformed ordinary Linear Regression in test performance ($R^2 = 0.8962$), implying that mild L2 regularization may stabilize coefficient estimation under reduced feature dimensionality. However, the improvement remains limited, reflecting the inherent information constraint imposed by a single predictor variable.

Tree-based ensemble models again demonstrated superior robustness. Random Forest achieved the highest test performance in this configuration ($R^2 = 0.9287$; MSE = 0.0022), followed by Gradient Boosting ($R^2 = 0.9096$). These results indicate that even when operating on a single feature, nonlinear partitioning strategies can recover additional structure from subtle variations in the RGB-B intensity distribution. Nevertheless, the decline in performance compared to the three- and nine-feature configurations confirms that higher-dimensional color representations encode complementary information not recoverable from RGB-B alone.

The Neural Network exhibited the weakest and most unstable behavior (test $R^2 = 0.8540$; CV std = 0.5592), with a marked discrepancy between training and cross-validation metrics. This instability reflects the limited representational benefit of deep architectures when only a single scalar input is available, further supporting the argument that neural models are not well-suited for low-dimensional structured datasets without significant complexity in feature space.

From a methodological standpoint, the one-feature experiment serves two critical purposes. First, it quantifies the predictive capacity of the most linearly correlated channel in isolation, effectively providing a calibration-style baseline. Second, it demonstrates that while RGB-B alone explains a substantial portion of variance, multi-channel configurations significantly enhance precision (lower MSE and MAE) and stability (lower CV variance). The progression from nine to four to three and finally one feature thus forms a structured reduction sequence, revealing how predictive fidelity degrades as color-space information is incrementally removed.

Overall, the results confirm that RGB-B constitutes a strong but insufficient standalone predictor for high-precision oxidant quantification. While acceptable accuracy may be achieved under minimalistic conditions, optimal performance---particularly in terms of error minimization and cross-validation stability---requires incorporating additional chromatic dimensions. This finding reinforces the rationale for multi-channel color-space integration in robust ML-based oxidant monitoring systems.

\subsubsection{Comparison of all models}

\begin{figure}[H]
    \centering
    \includegraphics[width=\linewidth]{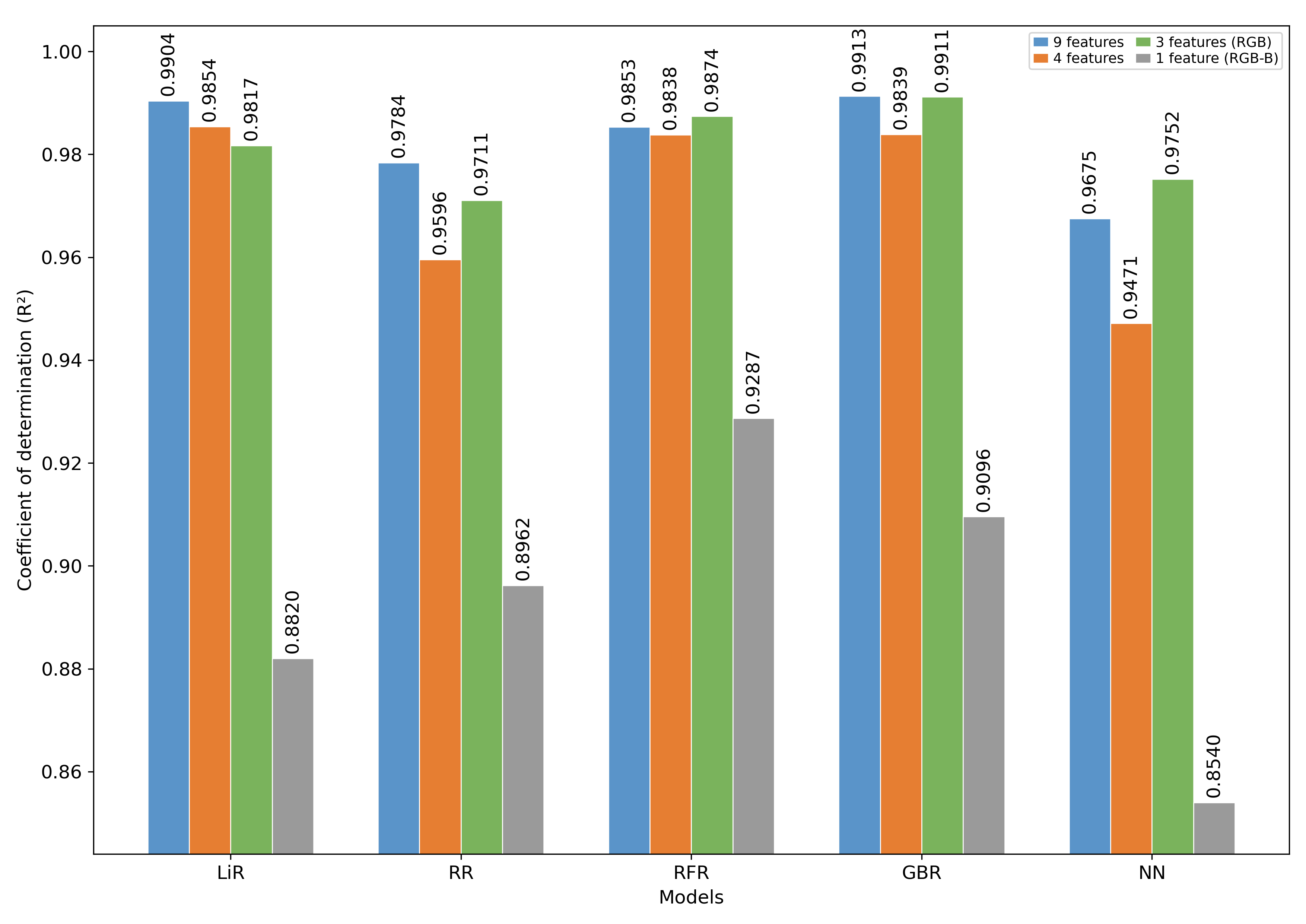}
    \caption{Comparison of the coefficient of determination ($R^2$) of ML models trained with nine vs. four features}
    \label{fig10}
\end{figure}

A direct performance comparison between models trained using nine-, four-, three-, and one-feature configurations provides deeper insight into the structural behavior of the predictive framework beyond the single-metric evaluation shown in Figure~\ref{fig10}. Rather than evaluating models solely under the full feature space, the progressive dimensionality reduction serves as a controlled reduction strategy to assess robustness, generalization, and sensitivity to feature sparsity.

Under the nine-feature configuration (RGB, HSV, Lab), both Linear Regression (LiR) and Gradient Boosting Regressor (GBR) achieved near-identical predictive performance ($R^2 \approx 0.99$), confirming that the dominant signal in the dataset is largely linear. This aligns with the strong univariate relationship observed between RGB-B and $[\text{Ox}]_{\text{tot}}$. However, comparable peak $R^2$ values do not necessarily imply identical structural robustness across modeling strategies.

When the feature set was reduced from nine to four statistically selected channels (RGB-B, HSV-S, Lab-a, Lab-b), Linear Regression exhibited a measurable decrease in test performance (0.9904 $\rightarrow$ 0.9854), whereas ensemble-based models (RFR and GBR) retained nearly stable predictive accuracy. This behavior supports findings by Schwendinger et al. (2024), who demonstrated that correlation-driven subset selection can stabilize linear estimators but may alter variance representation when multicollinearity is reduced \cite{Schwendinger2024AutomatedModels}. In contrast, tree-based ensembles dynamically prioritize informative splits and remain less sensitive to moderate dimensionality reduction \cite{Liu2025RandomizationForests,Li2025ErrorsThem}.

Further restriction to three RGB-only features revealed more pronounced differentiation. Linear Regression maintained high explanatory power ($R^2 = 0.9817$) but exhibited a clearer deviation from the nine-feature baseline. Gradient Boosting, however, preserved near-baseline performance ($R^2 \approx 0.991$), suggesting that nonlinear partitioning of RGB space captures chromatic interactions not fully represented by a single global linear mapping.

The single-feature configuration (RGB-B only) provides the most critical comparison. While Linear Regression remained operational ($R^2 = 0.8820$), its explanatory capacity declined substantially relative to higher-dimensional setups. Random Forest achieved a markedly higher $R^2$ (0.9287), indicating that nonlinear ensemble partitioning extracts additional structural information from localized intensity variations even when operating on a single predictor. This observation is consistent with theoretical analyses showing that randomization in tree ensembles can reduce both bias and variance under constrained feature conditions \cite{Liu2025RandomizationForests}.

Neural Network models exhibited comparatively unstable behavior, particularly under reduced feature conditions, reflected by elevated cross-validation variance. This aligns with established findings that deep architectures often underperform on structured, low-dimensional tabular datasets without extensive tuning \cite{Shwartz-Ziv2021TabularNeed}.

Across all configurations, cross-validation stability further differentiates model behavior. Ensemble methods maintained consistently low fold-to-fold variability, whereas Neural Networks demonstrated substantial dispersion under feature-restricted scenarios. This supports the interpretation that robustness under dimensional stress---not only peak predictive accuracy---is a key criterion for model selection. The importance of feature-driven stability and controlled dimensionality reduction has also been emphasized in ridge regression theory and provable feature selection guarantees \cite{Paul2015FeatureGuarantees}.

Collectively, these findings demonstrate that while a simple linear calibration based solely on RGB-B is mathematically feasible, multi-feature and ensemble-based approaches provide: (i) improved resilience under feature reduction, (ii) lower sensitivity to sampling variability, and (iii) enhanced predictive stability under information-constrained scenarios.

Therefore, the contribution of the ML framework in this study does not rest solely on outperforming linear regression in idealized conditions, but on demonstrating structural robustness and generalization capacity across controlled reductions in feature dimensionality. The nine-to-one feature progression functions as a methodological stress test, revealing how predictive fidelity degrades as color-space information is incrementally removed and highlighting the comparative resilience of ensemble learning approaches within the CBCA system.

\subsection{Step-wise Evaluation of Image Processing Stages on Predictive Performance}

Step-wise $R^2$ analysis was performed to quantify the incremental impact of each image-processing stage on predictive accuracy and to identify operations that enhance or impair model generalization. Rather than assessing only the final optimized pipeline, prediction performance was monitored across sequential transformations from raw image acquisition to multi–color-space feature extraction. At each stage, extracted color features were used to train and evaluate five machine learning models (GBR, LiR, NN, RFR, and RR) under identical preprocessing conditions (MinMax normalization, 80/20 train–test split, five-fold cross-validation), and test $R^2$ values were compared step-wise.

\begin{figure}[H]
    \centering
    \includegraphics[width=1\linewidth]{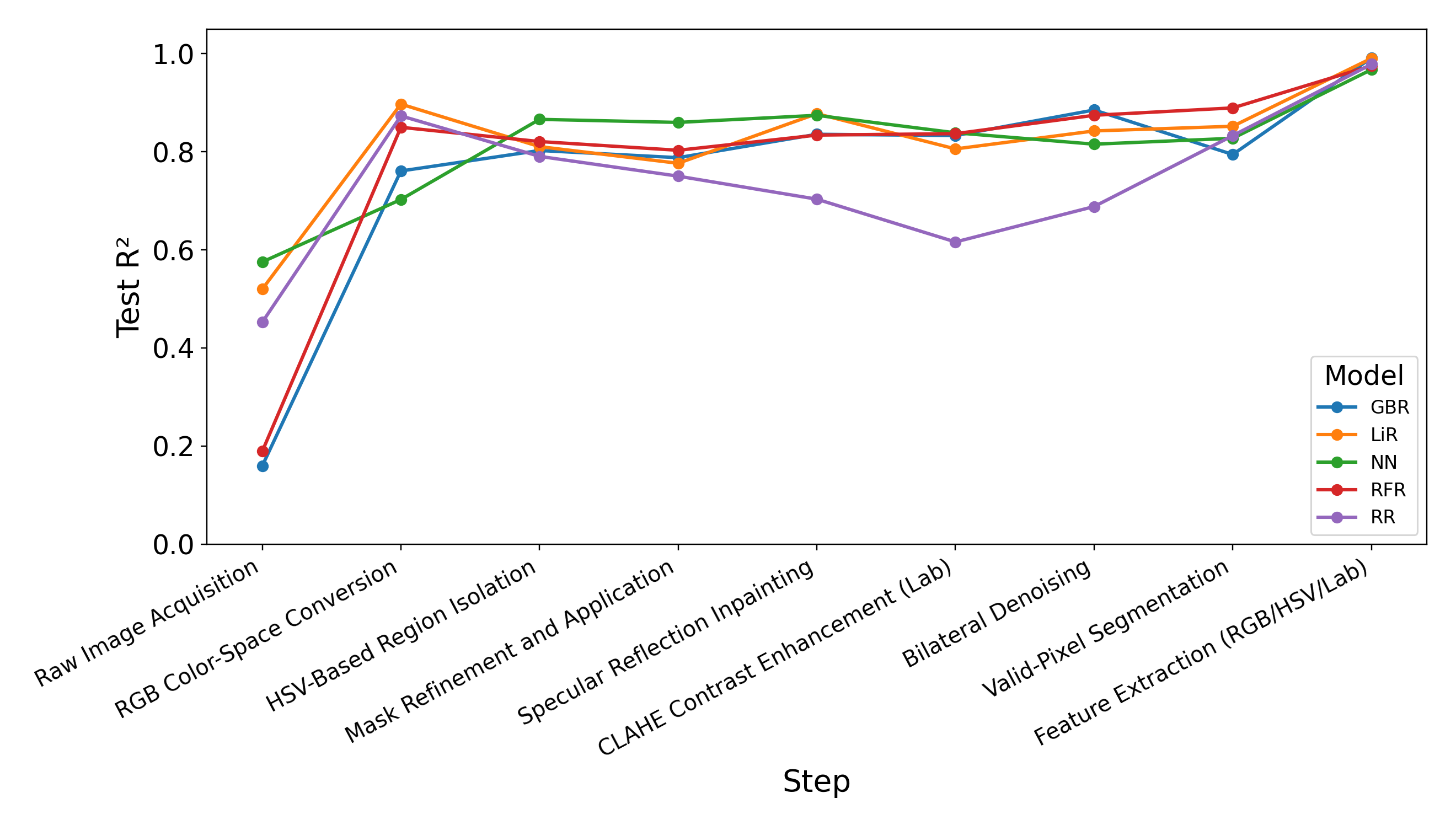}
    \caption{Step-wise comparison of test $R^2$ values for five machine learning models (GBR, LiR, NN, RFR, RR) across sequential image-processing stages. The analysis quantifies the incremental effect of each preprocessing operation on predictive performance and model generalization.}
    \label{fig:stepr2}
\end{figure}

The results demonstrate a substantial improvement immediately after color-space conversion, where $R^2$ values increased from 0.16--0.58 (raw images) to 0.70--0.90 across models, indicating that chromatic transformation captures the dominant concentration-related signal (Figure \ref{fig:stepr2}). Subsequent masking and reflection correction yielded moderate yet consistent gains, particularly for GBR and RFR, suggesting improved signal isolation. In contrast, CLAHE in Lab space and bilateral filtering produced model-dependent effects; notably, RR performance decreased to $R^2 = 0.62$ after CLAHE, implying potential over-enhancement or distortion of linear relationships. 

The final feature extraction stage resulted in uniformly high performance ($R^2 = 0.967$--0.991), confirming that cumulative preprocessing effectively maximizes predictive information. Overall, the analysis reveals that early color-space transformation and targeted masking contribute most strongly to generalization, whereas aggressive contrast enhancement may not universally benefit linear models. This evidence-based evaluation supports selective retention of high-impact steps while reconsidering operations that introduce instability or marginal gains.

\subsection{Validation of the proposed CBCA method with measured data}

\begin{figure}[H]
    \centering
    \includegraphics[width=0.9\linewidth]{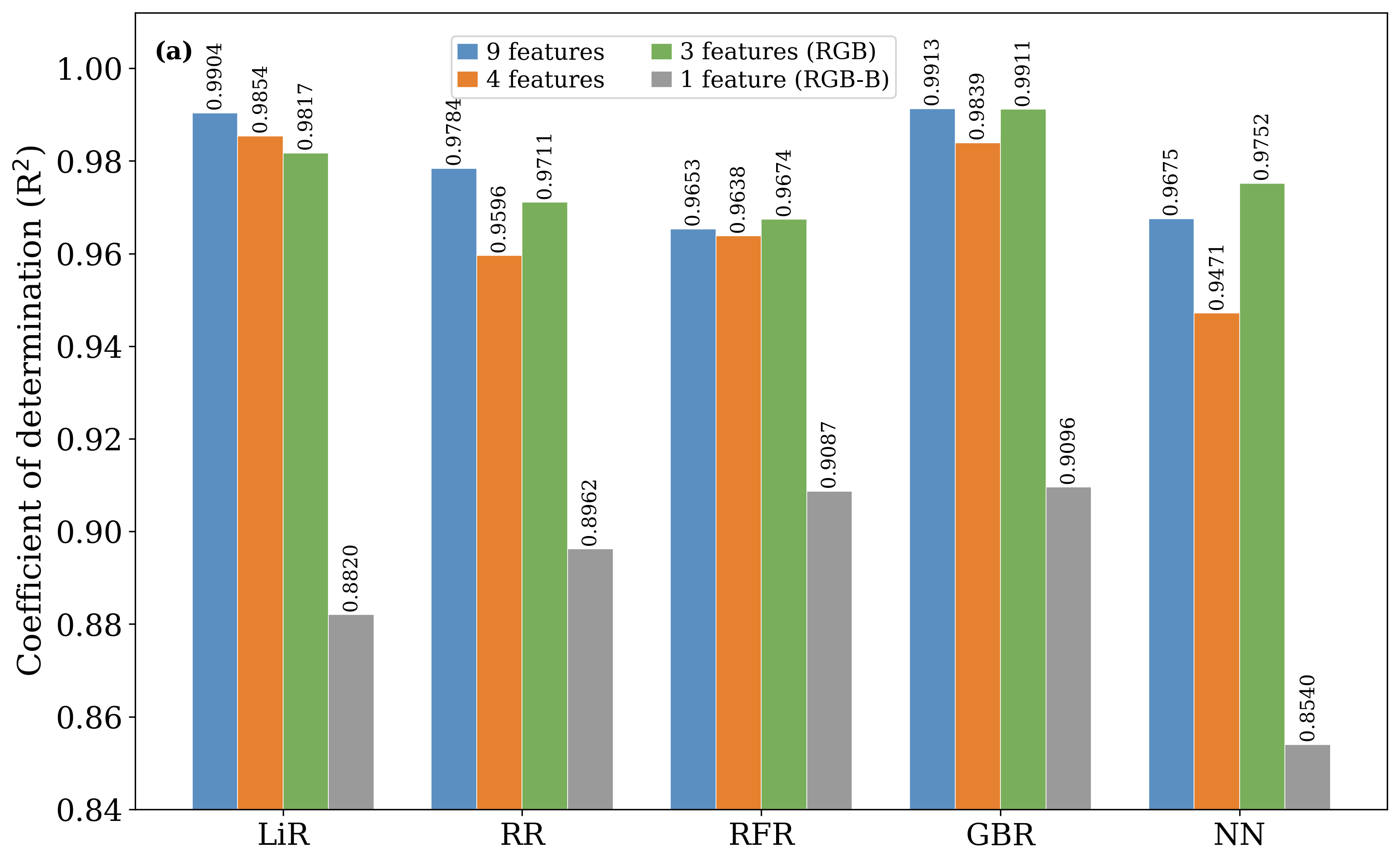}\\[0.75em]
    \includegraphics[width=0.9\linewidth]{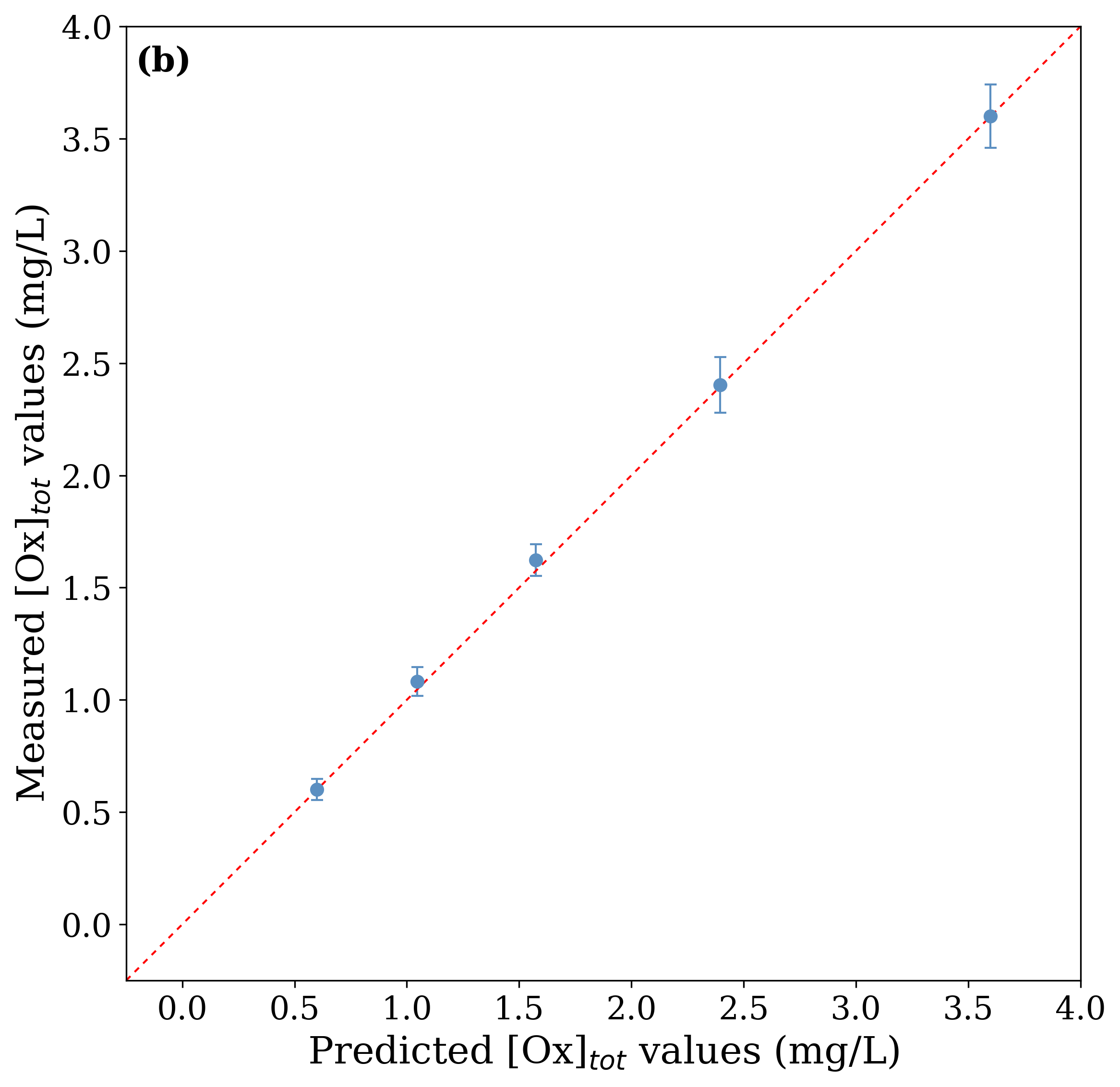}

    \caption{The results of the method validation: (a) comparison of the correlation coefficients ($R^2$) of the five ML algorithms with different number of features, (b) comparison of the measured and predicted concentrations of the GBR algorithm with four features}
    \label{fig11}
\end{figure}

To re-evaluate the analytical validity of the proposed CBCA framework, model predictions were systematically compared with experimentally determined total oxidant concentrations $[\text{Ox}]_{\text{tot}}$ obtained via the KI method using four distinct feature configurations as nine-, four-, three-, and single-feature inputs (Figure~\ref{fig11}).

To evaluate the analytical validity of the proposed CBCA method, model predictions were compared against experimentally determined total oxidant concentrations $[\text{Ox}]_{\text{tot}}$ obtained via the KI method. As shown in Figure~11a, all five machine learning models maintained high predictive performance across multi-feature datasets, with test $R^2$ values generally exceeding 0.96 and reaching up to 0.9913 for the GBR model under the nine-feature configuration, indicating that the CBCA approach provides analytically reliable predictions within experimental confidence limits.

Among the evaluated configurations, the nine-feature dataset yielded the highest overall predictive accuracy, where the GBR model achieved the best performance ($R^2 = 0.9913$), followed closely by LiR ($R^2 = 0.9904$) and RFR ($R^2 = 0.9653$). These results confirm that incorporating the full RGB, HSV, and Lab color-space information enables a more comprehensive representation of chromatic variations associated with oxidant concentration changes.

Among these, the RFR and GBR models stood out as the most stable performers throughout the validation process. When dimensionality was reduced from nine to four statistically selected features (RGB-B, HSV-S, Lab-a, and Lab-b), GBR maintained a high test $R^2$ of 0.9839, while RFR preserved a value of 0.9638, demonstrating only marginal performance reductions despite substantial input compression. This suggests a high degree of robustness and generalization, especially when operating with the four most statistically significant features: RGB-B, HSV-S, Lab-b, and Lab-a, selected based on correlation coefficients.

Figure~11a summarizes $R^2$ comparisons for all models before and after feature reduction. Dimensionality reduction produced model-dependent responses rather than uniform stability across all algorithms. While ensemble-based models (RFR and GBR) maintained consistently high predictive accuracy across all multi-feature configurations, simpler regression models such as RR and NN exhibited more noticeable performance decreases, particularly when restricted to four features. Although simpler models like Linear Regression (LiR) and Ridge Regression (RR) experienced modest performance drops (0.5\% and 2.81\%, respectively), their predictions remained within acceptable analytical margins, reaffirming the consistency of the CBCA method across model types. This is in line with prior findings that both RFR and GBR models exhibit minimal sensitivity to data-driven randomness and consistently deliver stable results across resampling procedures \cite{Raste2022QuantifyingAlgorithms}.

Notably, the three-feature RGB-only configuration yielded predictive accuracies nearly equivalent to the nine-feature case for ensemble-based models. The GBR model achieved a test $R^2$ of 0.9911 using only RGB inputs, indicating that the essential chromatic information required for oxidant quantification is largely preserved within the primary RGB color space under controlled imaging conditions.

In contrast, the single-feature RGB-B configuration resulted in a pronounced reduction in predictive accuracy across all models, with test $R^2$ values ranging from 0.8540 (NN) to 0.9096 (GBR). This outcome confirms that reliance on a single spectral channel is insufficient to fully represent the nonlinear colorimetric response associated with total oxidant formation, thereby establishing a lower bound for acceptable feature dimensionality in CBCA-based quantification.

To further validate the models under experimental conditions, additional tests were performed across five distinct plasma exposure durations (2, 5, 8, 12, and 16 minutes), corresponding to an $[\text{Ox}]_{\text{tot}}$ range of 0.66--3.50~mg/L. During each interval, video-recorded solution images were processed to extract color features and predict oxidant concentrations. Figure~11b compares the GBR\textquoteright s predictions with the experimental measurements using the nine-feature configuration. Each data point represents experiments that were done in our laboratory.

A refined comparison between predicted and measured concentrations using the optimal GBR configuration further confirmed the strong quantitative agreement between CBCA-derived outputs and KI-based measurements. As illustrated in Figure~11b, predicted concentrations closely follow the 1:1 reference line across the full working range (0.05--0.64~mg/L), demonstrating minimal systematic deviation and a highly linear response.

Analysis of the paired prediction dataset indicates that residuals remain symmetrically distributed around the reference line, with no observable heteroscedastic behaviour across concentration levels. Minor deviations observed in mid-range concentrations remain within the expected analytical uncertainty of manual titration procedures, confirming the quantitative reliability of the CBCA approach.

The alignment of predictions along the red 1:1 reference line and the narrow, symmetric error bars confirm a strong agreement between CBCA-derived outputs and titration measurements. While all models demonstrated valid and reliable performance, the RFR and GBR models provided the most consistent and stable predictions across both validation and experimental testing stages, reinforcing their suitability for deployment in automated oxidant quantification applications.

Importantly, ensemble-based learning algorithms, particularly GBR, maintain predictive stability even under feature-restricted conditions, highlighting the method’s robustness for practical implementation in real-time and resource-limited analytical systems.

This graphical validation demonstrates that the CBCA method ensures both precision and reliability, achieving predictive accuracy that remains within the experimental uncertainty limits of the KI method. As shown in Figure~11b, the prediction error margins of the GBR are closely aligned with those observed in manual titration, suggesting that the model\textquoteright s output variance lies well within the expected range of analytical variability. Such proximity between predicted and measured values reaffirms the CBCA system\textquoteright s suitability as a robust, operator-independent alternative to traditional titration techniques.

More importantly, the consistently high predictive accuracy observed across all evaluated models underscores the overall robustness of the CBCA approach. While all models operated within valid analytical tolerance, the ensemble-based algorithms, particularly RFR and GBR, distinguished themselves by producing the most stable and reliable results across both full and reduced feature configurations. Their resilience to feature reduction without significant loss of accuracy emphasizes their adaptability and reinforces their value for practical deployment.

Overall, the CBCA methodology maintains strong analytical fidelity across varying feature dimensionalities and modelling conditions. The demonstrated agreement between machine learning predictions and reference titration measurements supports the integration of CBCA-based quantification into automated, image-driven analytical platforms for rapid oxidant monitoring.

This study contributes to the growing body of research on artificial intelligence (AI) in analytical chemistry by introducing a novel approach for the rapid and accurate determination of $[\text{Ox}]_{\text{tot}}$ values. The results demonstrate the capability of ML-based methodologies to model and interpret heterogeneous datasets, thereby supporting the optimization of analytical processes \cite{CardosoRial2024AIDirections}. In this context, our work expands the current applications of AI in analytical chemistry and opens new avenues for future research in critical areas such as spectral interpretation, chromatographic condition optimization, and ensuring data integrity.

\section{Conclusion}
This study developed and validated a color-based computer analysis (CBCA) framework for quantifying total oxidant concentration ($[\text{Ox}]_{\text{tot}}$) in KI solutions exposed to non-thermal plasma (NTP), aiming to reduce operator-dependent variability inherent to conventional iodometric titration workflows. Using the Genesis controlled-imaging chamber and video-based acquisition, a sequential image-processing pipeline (ROI segmentation, masking, artifact correction, contrast/noise handling, and feature extraction) enabled reproducible derivation of RGB, HSV, and Lab descriptors from a continuous bubble-column oxidation environment.

Statistical and regression analyses demonstrated that concentration-linked chromatic response is dominated by a small subset of channels: HSV-S and Lab-b exhibited the strongest positive association with $[\text{Ox}]_{\text{tot}}$, while RGB-B showed the strongest inverse linear relationship, collectively supporting their role as primary indicators of KI oxidation progress. Importantly, the step-wise evaluation across preprocessing stages showed that early color-space transformation and targeted masking/reflection correction provided the largest gains in predictive performance, whereas aggressive enhancement operations (e.g., CLAHE) produced model-dependent effects and may distort linear relationships in certain estimators.

Five machine learning models (LiR, RR, RFR, GBR, and NN) were trained and evaluated under a controlled reduction sequence (nine-, four-, three-, and single-feature configurations) using identical scaling, data splitting, and cross-validation procedures. Across the full nine-feature representation, LiR and GBR achieved near-ceiling accuracy ($R^2 \approx 0.99$), indicating that a substantial component of the signal is explainable by simple linear structure under standardized imaging conditions. However, the reduction experiments clarified the added value of the ML framework beyond a single-channel calibration: ensemble learners (especially GBR and RFR) consistently preserved predictive fidelity under feature restriction (including RGB-only inputs), while RR and NN showed greater sensitivity and higher fold-to-fold variability, highlighting robustness differences that are not captured by peak $R^2$ values alone. When reduced to one feature (RGB-B), performance declined across all models, establishing a practical lower bound and confirming that multi-channel information is required for high-precision quantification (lower error and improved stability).

Independent method validation experiments further confirmed strong agreement between CBCA predictions and KI-based titration measurements, with predicted-versus-measured trends closely following the 1:1 reference behavior and remaining within expected analytical uncertainty. Overall, the CBCA system demonstrates an operator-independent, scalable pathway for near real-time oxidant monitoring in plasma-treated KI solutions, with ensemble-based models providing the most reliable performance under realistic constraints on feature dimensionality and computational resources.

Future work should (i) expand validation under broader operational variability (lighting drift, camera settings, matrix effects, and different oxidant chemistries), (ii) formalize uncertainty propagation from imaging to concentration estimates, and (iii) integrate the pipeline into closed-loop or continuous monitoring architectures for plasma-assisted oxidation processes and related colorimetric reaction systems.

\section*{Acknowledgment}
The authors gratefully acknowledge the financial support provided by the Scientific and Technological Research Council of Turkey (T\"{U}B\.{I}TAK) under project number 122Y089.

\section*{Declarations}
\noindent\textbf{Supplementary Material:} The supplementary materials were shared on GitHub (\url{https://github.com/mirkanemirsancak/colorbasedcomputeranalysis}) which contains the code and data for image processing-based machine learning.

\noindent\textbf{Funding:} This study is part of a research project (Grant No. 122Y089) financially supported by the Scientific and Technological Research Council of Turkey (T\"{U}B\.{I}TAK).

\noindent\textbf{Conflicts of interest:} All authors certify that they have no affiliations with or involvement in any organization or entity with any financial interest or non-financial interest in the subject matter or materials discussed in this manuscript.

\noindent\textbf{Authors' contributions:} All authors contributed to the study conception and design. Material preparation, data collection, and analysis were performed by M.E.\ Sancak. The first draft of the manuscript was written by M.E.\ Sancak and U.D.\ Keris-Sen, and U.\ Sen revised previous versions of the manuscript. All authors read and approved the final manuscript.

\noindent\textbf{Ethics approval:} Not applicable.

\noindent\textbf{Consent to participate:} Not applicable.

\noindent\textbf{Consent for publication:} Not applicable.

\clearpage
\bibliographystyle{achemso}
\bibliography{references}

\end{document}